\providecommand{\version}{}
\documentclass[\version]{aa}

\usepackage{amsmath}
\usepackage[english]{babel}
\usepackage{booktabs}
\usepackage{enumerate}
\usepackage[T1]{fontenc}
\usepackage{hyperref}
\usepackage{graphicx}
\usepackage[utf8]{inputenc}
\usepackage{lipsum}
\usepackage{longtable}
\usepackage{txfonts}
\usepackage{xcolor}

\makeatletter
\renewcommand*\aa@pageof{, page \thepage{} of~\pageref*{LastPage}}
\makeatother

\newcommand{\NumberRedshiftLowObscurationAll}{{2\,642}}
\newcommand{\Naegis}{{252}}
\newcommand{\Nbootes}{{1\,992}}
\newcommand{\Ncclccosmos}{{753}}
\newcommand{\Ncdfsfourms}{{154}}
\newcommand{\Ncdfssevenms}{{110}}
\newcommand{\Ncdfn}{{61}}
\newcommand{\Nxmmxxl}{{2\,129}}
\newcommand{\Nxuds}{{114}}
\newcommand{\Ntotal}{{5\,565}}

\newcommand{\NumberRedshiftHighObscurationAll}{{2\,923}}

\newcommand{\BiasAllevatoRedshiftLowObscurationObscured}{{{1.43}_{-0.09}^{+0.11}}}

\newcommand{\BiasAllevatoRedshiftLowObscurationCTK}{{{1.76}_{-0.19}^{+0.18}}}

\newcommand{\LogMhaloAllevatoRedshiftLowObscurationUnobscured}{{{12.74}_{-0.11}^{+0.09}}}

\newcommand{\LogMhaloAllevatoRedshiftHighObscurationUnobscured}{{{12.41}_{-0.12}^{+0.13}}}

\newcommand{\LogMhaloAllevatoRedshiftLowObscurationCTK}{{{12.98}_{-0.22}^{+0.17}}}

\newcommand{\LogMhaloAllevatoRedshiftHighObscurationCTK}{{{12.28}_{-0.19}^{+0.13}}}

\newcommand{\LogMhaloAllevatoRedshiftLowObscurationCTKApprox}{{{13.0}}}

\newcommand{\LogMhaloAllevatoRedshiftHighObscurationCTKApprox}{{{12.3}}}

\begin{document}

\author{
A. Viitanen \inst{1}
\and V. Allevato \inst{2,3,1}
\and A. Finoguenov \inst{1}
\and F. Shankar \inst{4}
\and R. Gilli \inst{5}
\and G. Lanzuisi \inst{5}
\and F. Vito \inst{5}
}

\authorrunning{A.~Viitanen et al.}

\institute{
Department of Physics, University of Helsinki, PO Box 64, FI-00014 Helsinki, Finland \\
\email{akke.viitanen@helsinki.fi}
\and INAF-Osservatorio astronomico di Capodimonte, Via Moiariello 16, I-30131 Naples, Italy
\and Scuola Normale Superiore, Piazza dei Cavalieri 7, I-56126 Pisa, Italy
\and School of Physics \& Astronomy, University of Southampton, Highfield, Southampton SO17 1BJ, UK
\and INAF-Osservatorio di Astrofisica e Scienza delle Spazio di Bologna, OAS, Via Gobetti 93/3, 40129 Bologna Italy
}

\date{Received dayofmonth dd, yyyy, accepted dayofmonth dd, yyyy}

\title{%
  Large-scale clustering of buried X-ray AGN\@: Trends in AGN obscuration
  and redshift evolution%
}

\titlerunning{Large-scale Clustering of Buried X-ray AGN}

\abstract%
{}
{%
  In order to test active galactic nucleus (AGN) unification and evolutionary
  models, we measured the AGN clustering properties as a function of AGN
  obscuration defined in terms of hydrogen column density, ${N_\mathrm{H}}$. In
  addition to measuring the clustering of unobscured (${N_\mathrm{H}} <
  10^{22}\,\mathrm{cm}^{-2}$) and moderately obscured ($10^{22} \leq
  {N_\mathrm{H}} < 10^{23.5}$) AGNs, we also targeted highly obscured sources
  (${N_\mathrm{H}}\geq 10^{23.5}$) up to redshifts of $z=3$.
}{%
  We have compiled one of the largest samples of X-ray-selected AGNs from a total of eight
  deep \textit{XMM}/\textit{Chandra} and multiwavelength surveys. We measured the
  clustering as a function of both AGN obscuration and redshift using the
  projected two-point correlation function, $w_{\mathrm{p}}(r_{\mathrm{p}})$. We
  modeled the large-scale clustering signal, measured the AGN bias, $b(z,
  {N_\mathrm{H}})$, and interpreted it in terms of the typical  AGN host
  dark matter halo, ${M_\mathrm{halo}}(z, {N_\mathrm{H}}$).
}{%
  We find no significant dependence of  AGN clustering on obscuration,
  suggesting similar typical masses of the hosting halos as a function of
  $N_\mathrm{H}$. This result matches expectations of AGN unification models,
  in which AGN obscuration depends mainly on the viewing angle of the
  obscuring torus. We measured, for the first time, the clustering of highly
  obscured AGNs and find that these objects reside in halos with typical mass
  $\log M_\mathrm{halo} = \LogMhaloAllevatoRedshiftLowObscurationCTK{}[h^{-1}
  M_\odot]$ ($\LogMhaloAllevatoRedshiftHighObscurationCTK{}$)
  at low $z \sim 0.7$ (high $z \sim 1.8$) redshifts.
  We find that irrespective of obscuration, an increase in AGN bias with
  redshift is slower than the expectation for a constant halo mass and
  instead follows the growth rate of halos, known as the passive evolution track.
  This implies that for those AGNs the clustering is mainly driven by the mass
  growth rate of the hosting halos and galaxies across cosmic time.
}{}{}


\maketitle

\section{Introduction}%
\label{sec:introduction}

Active galactic nucleus (AGN) unified models have been a central pillar of AGN
phenomenology for nearly four decades. The model states that the central
engines and environments of all AGNs are essentially the same. The black hole
(BH) at the galaxy center is surrounded by an accretion disk, and its growth
takes place behind a dusty, clumpy, obscuring torus \citep[e.g.,][]{antonucci93,
urry95, netzer15}. There is substantial evidence that this orientation-based
unification model is broadly applicable to local AGNs, whose nuclear regions
can be studied in detail \citep[e.g.,][]{honig13,tristram14}. However, it is now
recognized that such a model is oversimplified and requires ad hoc adjustments to
account for the wide range of AGN obscuration properties seen at different
redshifts and luminosities \citep[e.g.,][]{merloni14}. For example, the
increasing fraction of obscured AGNs toward high redshifts seems to indicate
that obscuration occurs not just in the torus but also on the scale of the host
galaxy \citep{lafranca05,ballantyne06,gilli2022A&A...666A..17G}, and that it
may be related to the overall galaxy evolution
\citep{treister06,hasinger08,ueda14,buchner15}.

Type 1 and 2 AGNs are postulated to be intrinsically the same objects in
orientation-based unification models, with Type 1, or broad-line, AGNs showing
broad emission lines (full width at half maximum $> 1000$ km/s)  and Type 2, or narrow-line, AGNs lacking
these broad emission features in their optical spectra. Historically, Type 2
AGNs have been described as an obscured version of Type 1 AGNs, with the broad-line-emitting region being hidden behind the partially opaque torus.

An additional way to classify AGNs based on obscuration properties is by
using the neutral gas column density, ${N_\mathrm{H}}$, along the line of sight,
as derived by X-ray spectral analysis or from the hardness ratio (HR\@), which is defined from X-ray
counts in hard ($\mathrm{H}$) and soft ($\mathrm{S}$) bands as $\mathrm{HR} =
(\mathrm{H} - \mathrm{S}) / (\mathrm{H} + \mathrm{S})$. Following this
approach, AGNs are classified as unobscured (${N_\mathrm{H}} < 10^{22}
\,\mathrm{cm}^{-2}$), obscured ($10^{22} < {N_\mathrm{H}} < 10^{24}$), Compton
thin (CTN), or highly obscured (${N_\mathrm{H}} > 10^{24}$) Compton thick
(CTK) AGNs\@. The optical type classification does not perfectly match with
the X-ray classification, as shown in \citet{merloni14} by using the
\textit{XMM}-COSMOS
\citep{hasinger2007ApJS..172...29H,cappelluti2009A&A...497..635C} AGN
catalog. Several studies have also used Wide-field Infrared Survey Explorer
\citep[WISE; ][]{wright2010AJ....140.1868W}
colors to define infrared-selected AGNs \citep{yan13} as obscured and
unobscured samples at $z<1.5$, by defining a mid-infrared-to-optical color cut
of ($r-W2$) $\sim$ 6 to separate these sources.

According to AGN unified models, unobscured Type 1 and obscured Type 2 AGNs
should have similar distributions in terms of redshift, luminosity, host galaxy
properties, and BH\ mass. In contrast, in the AGN evolutionary scenario,
obscured Type 2 AGNs may represent an earlier evolutionary phase compared to
unobscured systems and may have different properties \citep{hopkins08}. For
instance, if the AGN activity is triggered by a sporadic gas supply, unobscured
and obscured phases may occur several times during the galaxy lifetime. The
corresponding duty cycle, and its relation to the environment, should produce
distinctive statistical properties for the two populations \citep{hickox11}.

The clustering of AGNs provides a unique way to test BH triggering scenarios and
understand the link between obscured and unobscured AGNs, through their connection with
hosting dark matter (DM) halos. The clustering of obscured AGNs has been studied in
the last few decades, and the results are still controversial
\citep[e.g.,][]{allevato14,donoso14}; observational biases might be
responsible for these inconsistent results. At $z \lesssim 1.2$, studies based
on auto- or cross-correlation function analyses report no significant differences
in the clustering of Type 1 or 2, and/or unobscured or obscured AGNs
\citep{%
ebrero09,%
coil09,%
gilli09,%
krumpe12,%
mountrichas12,%
geach13,%
mendez16,%
jiang16,%
krumpe18,%
powell18%
}.
However, in these works AGNs are classified as unobscured or obscured sources based on
different methods, for example\ the HR \citep{ebrero09,coil09},
$N_\mathrm{H}$ \citep{cappelluti10}, and WISE colors \citep{mendez16}, or based on
the presence or lack of broad emission lines in the AGN spectra
\citep{gilli09,krumpe12}.

At intermediate redshifts, $z \sim 1$, based on X-ray AGNs and mid-infrared-selected
quasars, it has been reported with various significance that obscured AGNs
cluster more strongly and reside in denser environments than their unobscured
counterparts
\citep{%
  hickox11,%
  elyiv12,%
  donoso14,%
  dipompeo14,%
  dipompeo15,%
  dipompeo16,%
  dipompeo17,%
  koutoulidis18%
}.
Also in this case, selection biases due to the different AGN type definitions
and/or obscuration cuts might affect the clustering results. Recently,
\citet{koutoulidis18} measured the clustering of 736 (720) unobscured
(obscured) X-ray-selected AGNs in five deep \textit{Chandra} fields over $0.6 < z < 1.4$,
and found obscured sources (${N_\mathrm{H}} > 10^{22} \,\mathrm{cm}^{-2}$ as
classified using the HR) to be slightly more clustered than
unobscured AGNs\@.

Furthermore, building samples of CTK AGNs is also difficult as high
sensitivities over large areas are needed to collect a sizable number of
objects. For this reason, there have been few attempts to estimate the clustering properties of
heavily obscured AGNs. Thanks to \textit{XMM-Newton} and
\textit{Chandra} surveys
\citep[e.g.,][]{%
tozzi2006A&A...451..457T,%
brightman2012MNRAS.423..702B,%
georgantopoulos2013A&A...555A..43G%
},
suitable samples of highly obscured AGNs in different surveys are now available
and can be combined and used for the first time for clustering analysis.

In this work we aim to measure the clustering properties as a function of
obscuration, defined in terms of the hydrogen column density, ${N_\mathrm{H}}$, of
X-ray-selected AGNs.\ In doing so, we have compiled one of the largest samples of AGNs from eight deep
\textit{XMM-Newton}/\textit{Chandra} surveys. For the first time, in addition to
measuring the clustering of unobscured (${N_\mathrm{H}} < 10^{22}
\mathrm{cm}^{-2}$) and moderately obscured AGNs ($10^{22} \le {N_\mathrm{H}} /
\mathrm{cm}^{-2} < 10^{23.5}$), we also specifically target the highly obscured
AGN ($\geq 10^{23.5}$) population, near the CTK limit. Wherever
applicable, we assume a flat $\Lambda$ cold dark matter (CDM) cosmology with $\Omega_\Lambda =
0.7$, $\Omega_\mathrm{m} = 0.3$, and $H_0 = 70
\,\mathrm{km}\,\mathrm{s}^{-1}\,\mathrm{Mpc}^{-1}$.

\section{Data}

\begin{figure*}[htbp]
  \centering
  \includegraphics[width=.49\linewidth]{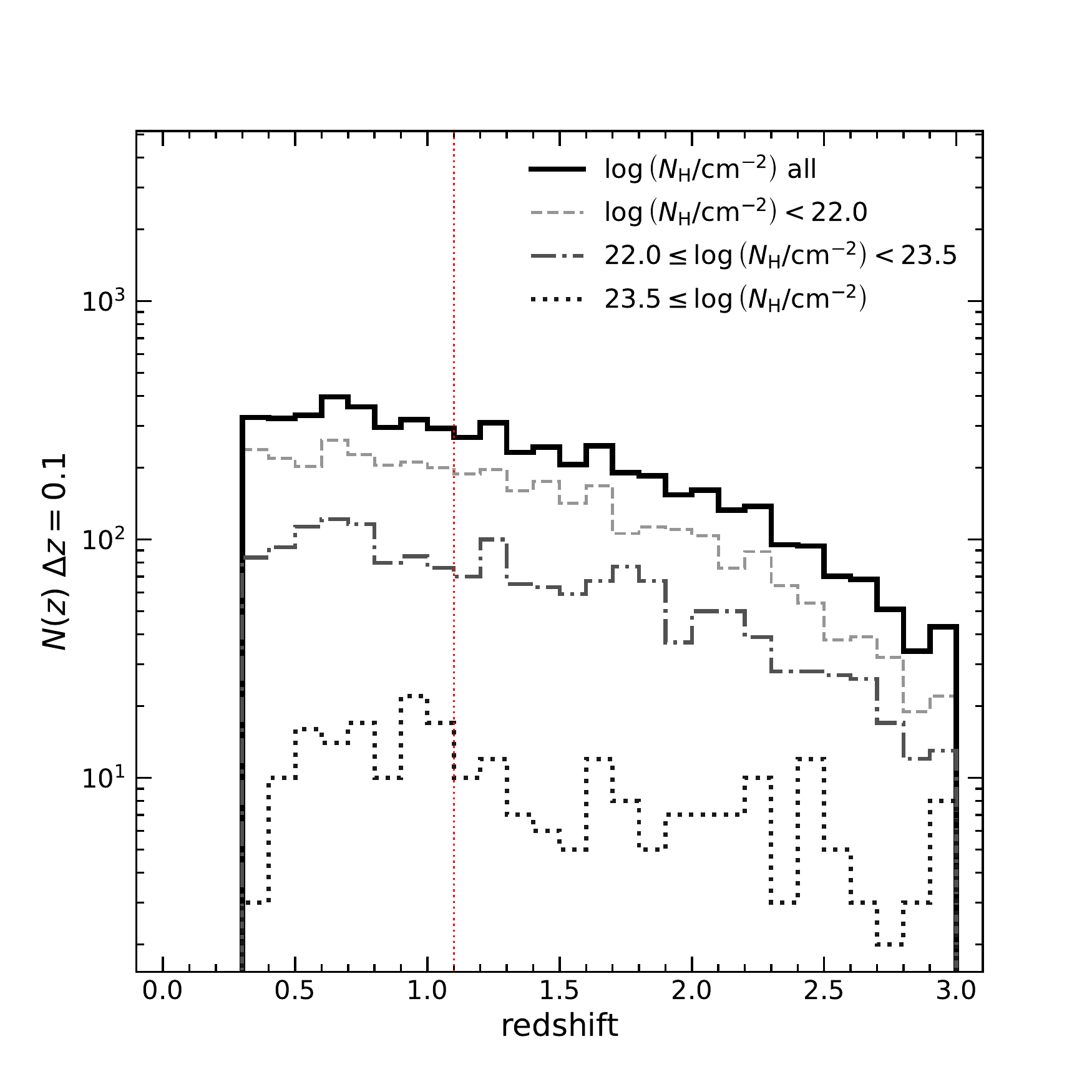}
  \includegraphics[width=.49\linewidth]{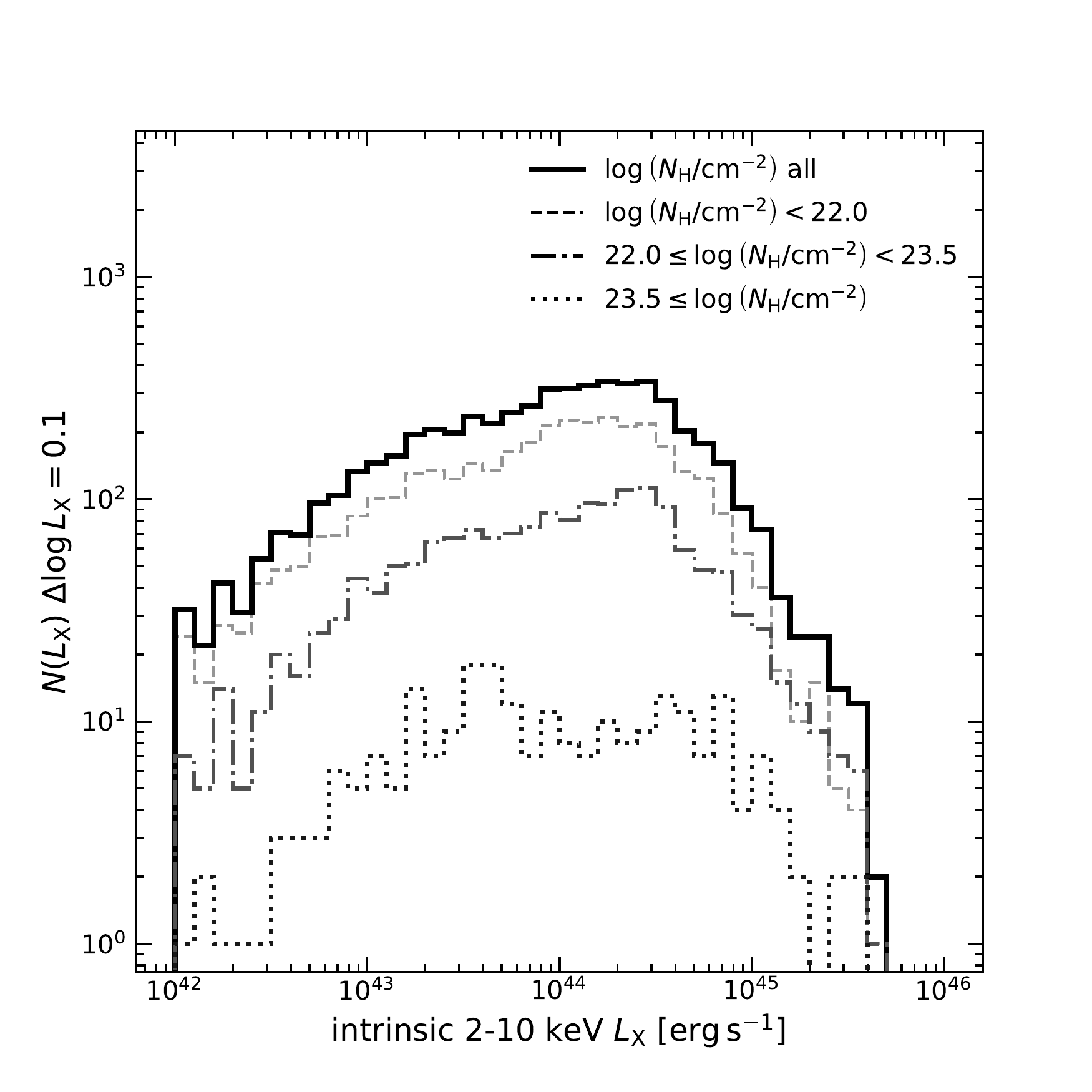}
  \includegraphics[width=.49\linewidth]{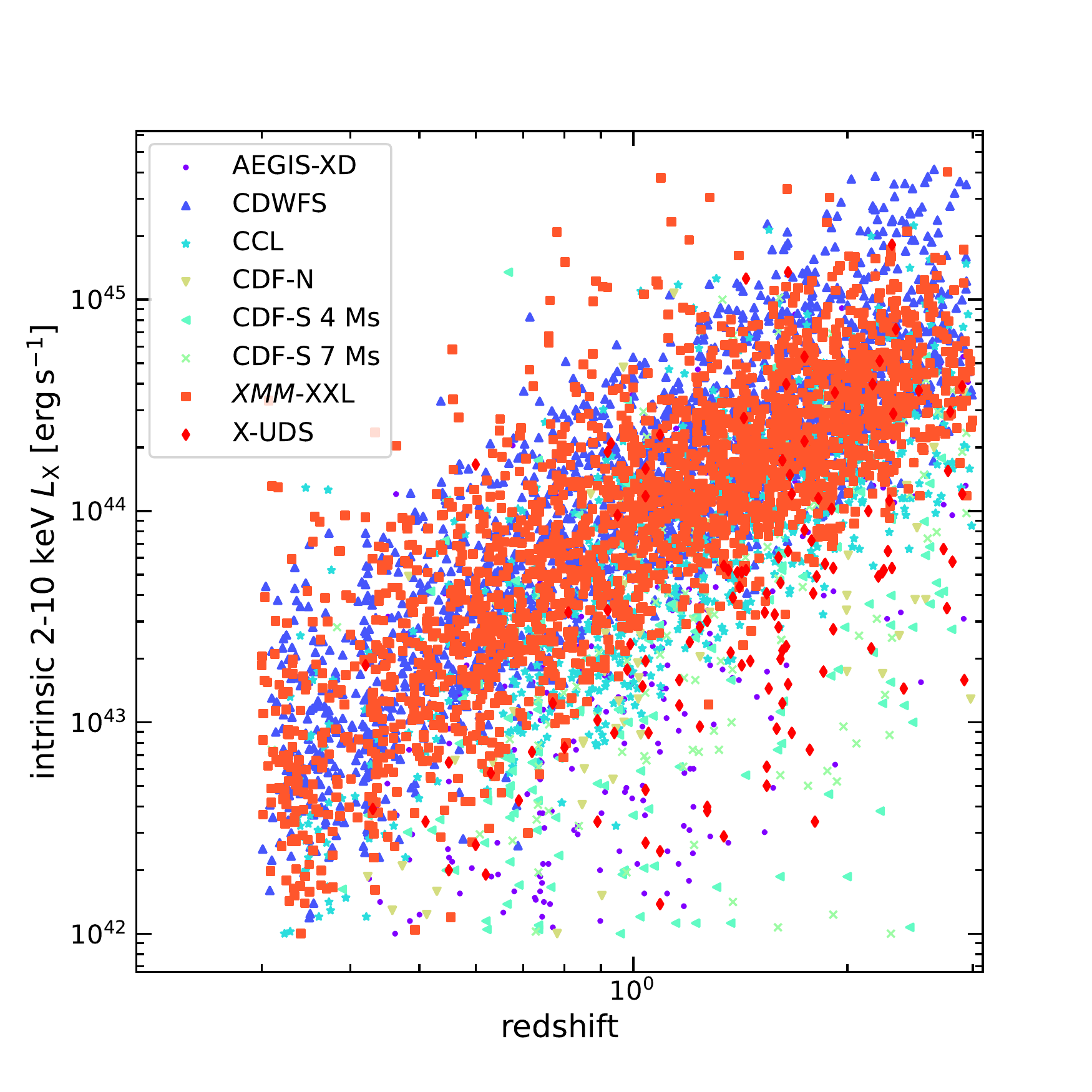}
  \caption{%
    Distribution of redshift, $z$, and intrinsic $2-10 \,\mathrm{keV}$ X-ray
    luminosity, $L_\mathrm{X}$, of the combined X-ray AGN sample with
    spectroscopic redshifts $0.3 \leq z < 3$ and $L_\mathrm{X} > 10^{42}
    \,\mathrm{erg}\,\mathrm{s}^{-1}$. The top-left panel shows the redshift
    distribution for the obscuration samples selected via hydrogen column
    density, ${N_\mathrm{H}}$, as indicated in the legend. The vertical red line
    indicates the low and high redshift bins. The top-right panel shows the
    distribution of $L_\mathrm{X}$ using the same line styles. The bottom panel
    shows the redshift against the intrinsic $2-10 \,\mathrm{keV}$ X-ray
    luminosity; different colors and markers indicate the various surveys
    used.
  }%
  \label{fig:zhist}
\end{figure*}

\begin{table*}[htbp]
\begin{center}
\begin{tabular}{llllll}
\toprule
\toprule
Survey & Reference & $N_\mathrm{specz}$ & $N_\mathrm{H} <    10^{22}$ & $             \geq 10^{22}$ & $             \geq 10^{23.5}$ \\
\midrule
AEGIS-XD       & \citet[][]{nandra15,brightman14}           &    $252$ &    $154$ &     $98$ &     $33$ \\
CDWFS          & \citet[][]{masini20}                       & $1\,992$ & $1\,002$ &    $990$ &     $24$ \\
CCL            & \citet[][]{civano16,marchesi16,lanzuisi18} &    $753$ &    $463$ &    $290$ &     $27$ \\
4 Ms CDF-S     & \citet[][]{xue11,rangel13,brightman14}     &    $154$ &     $90$ &     $64$ &     $14$ \\
7 Ms CDF-S     & \citet[][]{luo17,li20}                     &    $110$ &      $-$ &    $110$ &     $64$ \\
CDF-N          & \citet[][]{xue16,li20}                     &     $61$ &      $-$ &     $61$ &     $37$ \\
\emph{XMM}-XXL & \citet[][]{pierre16,liu16}                 & $2\,129$ & $1\,910$ &    $219$ &     $19$ \\
X-UDS          & \citet[][]{kocevski18}                     &    $114$ &     $41$ &     $73$ &     $23$ \\
\midrule
Total &  &  5565 &  3660 &  1905 &   241 \\
\bottomrule
\bottomrule
\end{tabular}
\caption{%
  Number of AGNs with secure spectroscopic redshifts (third
column) per survey for the eight surveys used in this study. The last three columns give
the number of AGNs above the indicated obscuration limit in cm$^{-2}$.
}%
\label{tab:field}
\end{center}
\end{table*}

In this work we study the clustering properties as a function of obscuration of
CTN (hydrogen column density ${N_\mathrm{H}} < 10^{23.5}
\,\mathrm{cm}^{-2}$) and highly obscured (${N_\mathrm{H}} \geq 10^{23.5}
\,\mathrm{cm}^{-2}$) X-ray-selected AGNs, compiling one of the largest AGN
data sets from deep \textit{XMM-Newton}/\textit{Chandra} X-ray and multiwavelength surveys
with available $N_H$ estimates. In detail, we combined a total of eight different
surveys: AEGIS-XD \citep[All-Wavelength Extended Groth Strip International Survey; ][]{nandra15},
the \textit{Chandra} Deep Wide-Field Survey (CDWFS) in the Boötes field \citep{masini20},
\textit{Chandra} Deep Field South (CDF-S) with 4 Ms and 7 Ms of exposure \citep{xue11,luo17},
\textit{Chandra} Deep Field-North \citep[CDF-N;][]{xue16}, \textit{Chandra} COSMOS
Legacy \citep[CCL;][]{marchesi16,civano16}, \textit{XMM}-XXL North \citep{liu16},
and X-UDS \citep[The Chandra Legacy Survey of the UKIDSS Ultra Deep Survey Field; ][]{kocevski18}.
These surveys cover a wide baseline in redshift, $0.04 < z <
5.3$, and were selected based on the availability of an AGN multiwavelength
catalog with spectroscopic redshifts and information on the AGN obscuration
${N_\mathrm{H}}$ derived from X-ray spectral fitting (seven out of the eight surveys) or the
X-ray HR (CDWFS). In the following sections, we summarize the
details of each survey and the corresponding AGN sample, including in the
analysis only AGNs with spectroscopic redshifts $0.3 \leq z < 3$ and intrinsic
$2-10\,\,\mathrm{keV}$ X-ray luminosities $>10^{42}
\,\mathrm{erg}\,\mathrm{s}^{-1}$.

\subsection{AEGIS-XD}%
\label{sec:aegis-xd}

AEGIS-XD \citep{nandra15} is a \textit{Chandra} program that focused on the central 0.29
deg$^2$ of the AEGIS field \citep[see][for the wider AEGIS-X]{laird09},
providing a nominal depth of 800 ks. The limiting X-ray fluxes are
$1.5 \times 10^{-16} \,\mathrm{erg}\,\mathrm{s}^{-1}\,\mathrm{cm}^{-2}$ (0.5$-$10 keV),
$3.3 \times 10^{-17} \,\mathrm{erg}\,\mathrm{s}^{-1}\,\mathrm{cm}^{-2}$ (0.5$-$2 keV),
$2.5 \times 10^{-16} \,\mathrm{erg}\,\mathrm{s}^{-1}\,\mathrm{cm}^{-2}$ (2$-$10 keV), and
$3.2 \times 10^{-16} \,\mathrm{erg}\,\mathrm{s}^{-1}\,\mathrm{cm}^{-2}$ (5$-$10 keV).
A total of 937 X-ray point sources are identified, 929 of which have multiwavelength
counterparts and 353 of which have spectroscopic redshifts. The X-ray spectral analysis
of the AGN sample (excluding 14 of the sources identified as stars\@) is
presented in \citet{brightman14}. They used the torus models set forth by
\citet{brightman11}, which are detailed in \citet[][Sect.~4.1]{brightman14}.
Of the 353 sources with spectroscopic redshifts, \Naegis{} fall within our selection limits based on AGN
redshifts and X-ray luminosities.\footnote{The \textit{Chandra} images, background, effective
exposure time, and sensitivity maps of AEGIS-XD are available at
\url{https://www.mpe.mpg.de/XraySurveys}}

\subsection{CCL}

The CCL survey \citep{civano16,marchesi16} is a 4.6 Ms
Chandra program focused on the 2.2 deg$^2$ COSMOS field \citep{scoville07}. The
CCL limiting fluxes are
$2.2 \times 10^{-16}\,\mathrm{erg}\,\mathrm{s}^{-1}\,\mathrm{cm}^{-2}$ (0.5$-$2 keV),
$1.5 \times 10^{-15}\,\mathrm{erg}\,\mathrm{s}^{-1}\,\mathrm{cm}^{-2}$ (2$-$10 keV), and
$8.9 \times 10^{-16}\,\mathrm{erg}\,\mathrm{s}^{-1}\,\mathrm{cm}^{-2}$ (0.5$-$10).
The overall spectroscopic completeness of the CCL is 53.6\%
\citep[$2\,151$ out of $4\,016$ sources;][]{marchesi16}.
We used a sample of $1\,949$ bright AGNs (with known spectroscopic and photometric
redshifts) with the obscuration $N_\mathrm{H}$ determined based on X-ray spectral analysis as
derived in \citet{marchesi16}. In addition, for a subset of $66$ CTK
AGN candidates, we used the obscuration derived by \citet[][]{lanzuisi18}.
Since a significant fraction of CCL sources have not been spectroscopically
followed up on in the outer regions of the CCL survey  \citep{marchesi16}, we
decided to focus only on the smaller central C-COSMOS region, which has a more
uniform spectroscopic coverage. The final number of CCL AGNs with
${L_\mathrm{X}} > 10^{42} \,\mathrm{erg}\,\mathrm{s}^{-1}$, robust obscuration
estimates from X-ray spectral analysis, and spectroscopic redshifts within $0.3
\leq z < 3$ is $\Ncclccosmos{}$ (with $405$ AGNs discarded from the outer CCL
survey).

\subsection{CDWFS}

The CDWFS \citep{masini20} is a new \textit{Chandra}
Legacy Survey in the Boötes field. \citet{masini20} analyzed all 281
\textit{Chandra} pointings in the $9.3 \,\mathrm{deg}^2$ Boötes field between 2003 and
2018 and present the X-ray point source catalog. The total exposure time is 3.4
Ms, and $6\,891$ X-ray point sources are detected. The limiting fluxes are $4.7
\times 10^{-16} \,\mathrm{erg}\,\mathrm{s}^{-1}\,\mathrm{cm}^{-2}$ (0.5$-$7.0
keV), $1.5 \times 10^{-16} \,\mathrm{erg}\,\mathrm{s}^{-1}\,\mathrm{cm}^{-2}$
(0.5$-$2.0 keV), and $9   \times 10^{-16}
\,\mathrm{erg}\,\mathrm{s}^{-1}\,\mathrm{cm}^{-2}$ (2$-$7 keV).
Spectroscopic and/or photometric redshifts are available for $94\%$ of the sources,
with a total of $2\,346$ spectroscopic redshifts available.
Obscuration- and absorption-corrected AGN X-ray luminosities were estimated
through a combination of assumed power-law spectra with $\Gamma =
1.8$ and a Bayesian estimate of the HR, namely,
$\mathrm{HR} = \mathrm{HR} (z, {N_\mathrm{H}})$ \citep[see Sect.~6.2 of][]{ricci17}. However,
the spectroscopic coverage of the CDWFS field is not homogeneous over the full
angular extent of the \textit{Chandra} pointings. In order to take this
effect into account and increase the spectroscopic redshift completeness of the survey, we
masked out the outer edges of the Boötes field that are not covered by AGES
\citep[AGN and Galaxy Evolution Survey; e.g.,][]{hickox09}. This leads to a reduction of $10\%$ in the size of
the original catalog (6268 sources out of 6891) presented in \citet{masini20}.
The final sample of CDWFS AGNs with spectroscopic redshifts contains \Nbootes{}
X-ray AGNs\@.

\subsection{CDF-S}

The CDF-S survey contains the deepest X-ray surveys ever
conducted. We combined two X-ray AGN samples from the field, one based on the 4
Ms CDF-S \citep{xue11,rangel13,brightman14} and the other targeting the
highly obscured AGNs (${N_\mathrm{H}} > 10^{23}\,\mathrm{cm}^{-2}$) in the 7 Ms
CDF-S \citep{luo17,li20}.

The limiting fluxes for the 4 Ms CDF-S are
$3.2 \times 10^{-17} \,\mathrm{erg}\,\mathrm{s}^{-1}\,\mathrm{cm}^{-2}$ (0.5$-$8 keV),
$9.1 \times 10^{-18} \,\mathrm{erg}\,\mathrm{s}^{-1}\,\mathrm{cm}^{-2}$ (0.5$-$2 keV), and
$5.5 \times 10^{-17} \,\mathrm{erg}\,\mathrm{s}^{-1}\,\mathrm{cm}^{-2}$ (2$-$8 keV)
\citep{xue11}. The corresponding limits for the 7 Ms CDF-S are
$1.9 \times 10^{-17} \,\mathrm{erg}\,\mathrm{s}^{-1}\,\mathrm{cm}^{-2}$ (0.5$-$7.0 keV),
$6.4 \times 10^{-18} \,\mathrm{erg}\,\mathrm{s}^{-1}\,\mathrm{cm}^{-2}$ (0.5$-$2.0 keV), and
$2.7 \times 10^{-17} \,\mathrm{erg}\,\mathrm{s}^{-1}\,\mathrm{cm}^{-2}$ (2$-$7 keV).
\citet{brightman14} conducted an X-ray spectral analysis of the 4 Ms CDF-S AGN
catalog of \citet{rangel13} following the same approach as for AEGIS-XD\@. For
the 7 Ms CDF-S, \citet{li19,li20} derived the obscuration ${N_\mathrm{H}}$
using the MYTorus code \citep{murphy09}, including several different torus
models. The total number of unique AGNs from the CDF-S 4 Ms and 7 Ms is 476, and
the final sample of AGNs with spectroscopic redshifts considered in this work is
\Ncdfsfourms{} (\Ncdfssevenms{}) AGNs from 4 Ms (7 Ms) CDF-S.\footnote{The corresponding
\textit{Chandra} maps for the 4 Ms (7 Ms) CDF-S are available at
\url{https://www.mpe.mpg.de/XraySurveys}
\citep[\url{http://www2.astro.psu.edu/users/niel/cdfs/cdfs-chandra.html};][]{luo17}.}

\subsection{CDF-N}

The CDF-N survey \citep{alexander03,xue16} is the second
deepest extragalactic $0.5-8.0 \,\mathrm{keV}$ survey ever conducted. It comprises 2 Ms of \textit{Chandra} exposure, covering 448 sq.\ arcmin. For 90\%
completeness, the flux limits are
$1.9 \times 10^{-15} \,\mathrm{erg}\,\mathrm{s}^{-1}\,\mathrm{cm}^{-2}$ (0.5$-$7 keV),
$6.0 \times 10^{-16} \,\mathrm{erg}\,\mathrm{s}^{-1}\,\mathrm{cm}^{-2}$ (0.5$-$2 keV), and
$2.7 \times 10^{-15} \,\mathrm{erg}\,\mathrm{s}^{-1}\,\mathrm{cm}^{-2}$ (2$-$7 keV).
The X-ray spectral analysis of obscured sources in 2 Ms CDF-N was conducted by
\citet{li20} in a similar way as in the previous section. For this work, we
included \Ncdfn{} obscured AGNs with spectroscopic redshifts in the analysis.
For CDF-N 2 Ms, the \textit{Chandra} exposure, background, and sensitivity maps
are available in \citet[][and references therein]{xue16}. We combined the
footprint of GOODS-N \citep[as reported by Fig.~1 of][]{li20} and CDF-N
\citep{xue16}.

\subsection{\textit{XMM}-XXL}

\textit{XMM}-XXL \citep{pierre16} is the largest \textit{XMM}-\textit{Newton} program
(as of 2016), with a total of 6.9 Ms of observations and two distinct 25
deg$^2$ fields, the northern (XXL-N) and the southern (XXL-S). The 90\%
limiting fluxes are
$4 \times 10^{-15} \,\mathrm{erg}\,\mathrm{s}^{-1}\,\mathrm{cm}^{-2}$ (0.5$-$2.0 keV) and
$2 \times 10^{-14} \,\mathrm{erg}\,\mathrm{s}^{-1}\,\mathrm{cm}^{-2}$ (2$-$10 keV)
for XXL-N\@. \citet{liu16} report on the X-ray spectral
properties of XXL-N, with a total of $2\,512$ analysed AGNs with reliable
X-ray spectra. The X-ray spectral analysis uses a combination of a torus model
\citep[][]{brightman11}, a reflection component \citep{nandra07}, and a soft
scattering component (see Sect.~4.1 of \citeauthor{liu16}). Further, in order to
account for the spectroscopic visibility mask of the optical follow-up, we
limited our analysis to the area covered by the five spectral plates in the
SDSS-III/BOSS ancillary programs \citep[see][]{menzel16,liu16}. The final
number of AGNs included in our analysis is \Nxmmxxl{}.

\subsection{X-UDS}

X-UDS \citep{kocevski18} is a $\sim 1.3$ Ms \textit{Chandra} program of the
Subaru-\textit{XMM} Deep/UKIDSS Ultra Deep Survey (UDS) field that covers an area
of $0.33$ deg$^2$. The X-UDS limiting fluxes are
$4.4 \times 10^{-16} \,\mathrm{erg}\,\mathrm{s}^{-1}\,\mathrm{cm}^{-2}$ (0.5$-$10 keV),
$1.4 \times 10^{-16} \,\mathrm{erg}\,\mathrm{s}^{-1}\,\mathrm{cm}^{-2}$ (0.5$-$2 keV), and
$6.5 \times 10^{-16} \,\mathrm{erg}\,\mathrm{s}^{-1}\,\mathrm{cm}^{-2}$ (2$-$10 keV).
The X-ray spectral analysis follows \citet[see
Sect.~\ref{sec:aegis-xd}]{brightman14}, using the torus model described
in~\cite{brightman11}. In order to select X-ray AGN counterparts that have
reliable spectroscopic redshifts, we limited the data we took from X-UDS to the
central CANDELS \citep[][]{grogin2011ApJS..197...35G} region. The final sample
of X-UDS AGNs with spectroscopic redshifts considered in this work contains
\Nxuds{}.\footnote{The X-UDS \textit{Chandra} maps are available at
\url{https://www.mpe.mpg.de/XraySurveys}.}

\subsection{Combined X-ray AGN sample}

The final combined AGN sample consists of a total of $\Ntotal{}$ AGNs with
spectroscopic redshifts $0.3 \leq z < 3.0$ (mean $z \sim 1.3$) and hydrogen
column densities, $N_\mathrm{H}$, derived through either X-ray spectral analysis or from the HR for CDWFS\@. The numbers of AGNs as a function of
different obscuration cuts in $N_\mathrm{H}$ are presented in
Table~\ref{tab:field}, while the distributions in redshift and intrinsic X-ray
luminosity ($2-10$ keV) are shown in Fig.~\ref{fig:zhist}.

For the analysis, we defined three AGN subsamples based on different cuts in
${N_\mathrm{H}}$, which we refer to as unobscured (${N_\mathrm{H}} < 10^{22}
\,\mathrm{cm}^{-2}$), moderately obscured ($10^{22} \leq {N_\mathrm{H}} /
\,\mathrm{cm}^{-2} < 10^{23.5}$), and highly obscured (${N_\mathrm{H}} \geq
10^{23.5}$) sources. It is worth noting that although unobscured, obscured, and
highly obscured AGNs are not matched in terms of X-ray luminosity, at each
redshift of interest the median $L_\mathrm{X}$ is similar among the different
subsamples.

In addition, in order to study the AGN clustering evolution, we defined two bins
in redshift. We divided the full AGN sample based on $0.3 \leq z < 1.1$ and
$1.1 \leq z < 3.0$. Both redshift ranges were designed to have a comparable
number of objects: $\NumberRedshiftLowObscurationAll{}$ and
$\NumberRedshiftHighObscurationAll{}$ for the low and high redshift ranges,
respectively.

We also note that the hydrogen column density estimates and the corresponding
classification for a given source may vary when using different methodologies
\citep[see, e.g.,][their Figs.~18-19]{brightman14} applied to different surveys
\citep[e.g.,][]{castello-mor13}. However, our work does not focus on single
sources and individual ${N_\mathrm{H}}$ measurements and instead provides a
statistical study of large samples of unobscured, moderately obscured, and highly
obscured AGNs\@. As shown in Sect.~\ref{sec:combinations}, we have verified that
our results are robust in the sense that removing individual surveys from the
analysis and slightly shifting the ${N_\mathrm{H}}$ cut does not make a
significant difference.

\section{Methodology}
\label{sec:methodology}

Here we describe the methodology and models used to quantify the AGN clustering
as a function of the obscuration in different redshift bins. In detail, we
measured the AGN clustering by using the projected two-point correlation
function, $w_{\mathrm{p}}(r_{\mathrm{p}})$ \citep{davis_peebles83}, which is
independent of redshift space distortions. The estimator requires the
construction of a random catalog that acts as an unclustered distribution of
sources.

\subsection{Random catalog}

We constructed random catalogs separately for each field included in the
analysis, assigning random redshifts extracted from the smoothed AGN redshift
distribution in the considered survey. We used a Gaussian smoothing
kernel with $\sigma_z = 0.3$ as a compromise between overfitting features and
smoothing out the distribution.
We then drew random coordinates in right ascension and declination,
discarding or keeping sources based on the inhomogeneous coverage of the
considered X-ray survey. The nonuniform coverage is taken into account
either by a sensitivity map (detection limit flux) or by exposure time. Using the
sensitivity map method set forth by \citet{georgakakis08}
\citep[e.g.,][]{allevato11,viitanen19}, a random flux was drawn from the flux
distribution of the AGN sample for each position, and the random source was kept
or discarded based on whether it exceeds the flux limit at that position. We
note that due to the survey flux limits, sampling the AGN distribution
underpredicts the number of sources with faint fluxes compared to, for example,\ drawing
fluxes directly from the AGN $\log N - \log S$. We verified that the
correlation function measurement is robust against undersampling the lower
fluxes by performing a test where fluxes are instead sampled from the $\log N -
\log S$ as derived by \citet[][]{luo17} for 7 Ms CDF-S AGNs. We find the
correlation function measurement remains unchanged within the errors.

The sensitivity map method was applied to AEGIS-XD, 4 Ms CDF-S, CDF-N, CDWFS,
COSMOS, and \textit{XMM}-XXL. For 7 Ms CDF-S and X-UDS, we decided to
discard random objects based on the effective exposure time at the sky position
in a linear manner. In detail, we normalized the exposure at the proposed
position to unity based on the maximum value of the exposure map and drew a
uniform random number between 0 and 1, which gives the probability of discarding the
random source at the position. We note that both these methods are preferable
to assigning random positions simply drawn from the distributions of
coordinates in the real data catalog directly, which discards the clustering
signal in the perpendicular direction \citep[e.g.,][]{gilli05,koutoulidis18}.
For each survey, we constructed a random sample $300$ times the size of the data
sample.

\subsection{Two-point correlation function estimator}

We measured the two-point correlation function, which measures the excess
probability of finding an AGN pair above random at a given physical separation of
$r_{\mathrm{p}}$ (perpendicular) and $\pi$ (parallel):\begin{equation}
  \mathrm{d}P_{12}
  = n^2 \left[ 1 + \xi(r_{\mathrm{p}}, \pi) \right] \mathrm{d}V_1 \mathrm{d}V_2,
\end{equation}
where $n$ is the mean AGN number density, and we estimate $\xi(r_{\mathrm{p}},
\pi)$ using the standard \citet{landy_szalay93} estimator in bins of
$r_{\mathrm{p}}$, $\pi$ as
\begin{equation}
  \xi(r_{\mathrm{p}}, \pi)
  = \frac{dd(r_{\mathrm{p}}, \pi) - 2dr(r_{\mathrm{p}}, \pi) + rr(r_{\mathrm{p}}, \pi)}
         {rr(r_{\mathrm{p}}, \pi)},
  \label{eq:xi2d}
\end{equation}
where $dd$, $dr$, and $rr$ correspond to the normalized number of data-data,
data-random, and random-random pairs. Then, we estimated the projected two-point
correlation function, $w_{\mathrm{p}}$, by integrating  along the line of sight
up to $\pi_{\max}$ to get rid of redshift space distortions
\citep{davis_peebles83}:
\begin{equation}
  w_{\mathrm{p}}(r_{\mathrm{p}})
  = 2 \int_0^{\pi_{\max}} \xi(r_{\mathrm{p}}, \pi) \mathrm{d}\pi.
  \label{eq:wprp}
\end{equation}
In detail, we estimated $w_{\mathrm{p}}(r_{\mathrm{p}})$ using eight logarithmic
bins for $r_{\mathrm{p}} = 0.1-100 \,{h}^{-1}\mathrm{Mpc}$, and with $\pi =
0-100\,\mathrm{Mpc}/h$ ($\Delta \pi = 1 \,{h}^{-1}\mathrm{Mpc}$) for the pairs.
In order to prevent integration over (potentially) empty bins, we first
integrated the pairs up to $\pi_{\max}$ along the line of sight before
estimating $w_{\mathrm{p}}(r_{\mathrm{p}})$. We find $\pi_{\max}
\sim 20 \,{h}^{-1}\mathrm{Mpc}$ to be a good compromise between maximizing the
clustering signal and introducing noise into the estimator.

\subsection{Large-scale bias and typical dark matter halo mass}

The AGN distribution is a biased tracer of the underlying DM
distribution. That is, the AGN projected two-point correlation function,
$w_{\mathrm{p}}(r_{\mathrm{p}})$, is related to the DM one,
$w_{\mathrm{p},\mathrm{DM}}(r_{\mathrm{p}}, z)$, at large scales, $r_\mathrm{p} \gtrsim
1\,\mathrm{Mpc}/h$,  via the large-scale bias, $b$:
\begin{equation}
  w_{\mathrm{p}}(r_{\mathrm{p}}) = b_{\mathrm{2-halo}}^2
  w_{\mathrm{p},\mathrm{DM}}(r_{\mathrm{p}}, z),
  \label{eq:bias}
\end{equation}
where $w_{\mathrm{p},\mathrm{DM}}$ is estimated using our adopted cosmology
and the $\Lambda$CDM power spectrum fitting formulae of
\citet{eisenstein1998ApJ...496..605E}. The standard approach used in previous
clustering studies \citep[e.g.,][]{hickox09} is to derive the large-scale bias
factor from Eq.~\ref{eq:bias} and associate a typical DM halo mass,
$b({M_\mathrm{halo}}, z)$, according to the ellipsoidal collapse model of
\citet{sheth01} and the numerical results of \citet{vandenbosch02}. However, in
this approach the large-scale structure growth and the evolution of the bias
factor with redshift are not properly considered. In order to include the
effect of the redshift dependence of the bias associated with our use of large redshift intervals as the structures grow over time, we estimated the bias
factor, $\overline{b}_\mathrm{A11}$, the redshift, $\overline{z}_\mathrm{A11}$,
and the halo mass, $M_{A11}$, by using the methodology described by Eqs. 20-24
of~\cite{allevato11}. These estimates are weighted by the growth function and
include only the AGN pairs that contribute to the AGN large-scale clustering.
Table~\ref{tab:allevato2011} shows the weighted bias,
$\overline{b}_\mathrm{A11}$, weighted redshift, $\overline{z}_\mathrm{A11}$, and
the corresponding DM halo mass, $M_{A11}$, for all the different subsamples
considered in this analysis. It is worth noting that
$\overline{b}_\mathrm{A11}$ is defined with respect to the DM correlation
function at $z=0$ ($w_{\mathrm{p},\mathrm{DM}}(r_{\mathrm{p}}, z=0)$) and thus
cannot be directly compared to the large-scale bias derived in
Eq.~\ref{eq:bias}. For this reason, following~\cite{allevato11}, we also
estimated the $b_\mathrm{A11}$ bias associated with $M_\mathrm{A11}$ according to
\citet{sheth01}. This bias factor can be seen as properly corrected for the
large-scale structure growth and the evolution of the bias factor within the
wide redshift bins considered. As shown in Table~\ref{tab:allevato2011},
$b_\mathrm{A11}$ is $\sim$ 10-20\% smaller than $b_\mathrm{2-halo}$ (see
Table~\ref{tab:allevato2011}).

We note that the typical DM halo mass derived from the large-scale bias
may not reflect the true distribution of AGN host halo masses. This is
especially the case if the underlying AGN host halo mass distribution spans a
range of halo masses in different large-scale environments. For example,
\citet{leauthaud15} find a skewed distribution of AGN host halo DM masses in
COSMOS, where the typical DM halo masses reported earlier varied between their
median (lower) and mean (higher) values. Also, using DM halos from large N-body
simulations and empirically motivated AGN models,
\citet{aird2021MNRAS.502.5962A} find that the {true} bias calculated
directly from the DM halos is systematically lower than the observationally
measured one. However, in their semi-analytic AGN model, \citet{oogi20} find
that there is little difference in the large-scale bias derived from the
projected two-point correlation function and the halo occupation distribution, which in turn is sensitive to the halo mass distribution. While
conclusions about the typical AGN environment based on large-scale bias alone
may be limited, the main focus of this work is to point out relative
differences in AGN populations in terms of redshift and obscuration, which is more robust against the aforementioned limitations.

\subsection{Uncertainty estimation}

In order to estimate the two-point correlation function errors, as well as the subsequent errors on the
large-scale bias and the typical DM halo mass, we used a bootstrap
resampling method.
In detail, we divided the combined AGN sample into non-overlapping regions based
on right ascension and the sine of declination. The number of regions defined for
each survey was selected so that the sky area and number of objects per region
are comparable, and the total number of regions ranges from 1 (pencil beams,
e.g.,\ 7 Ms CDF-S) to 36 ($\emph{XMM}-XXL$). Then, we
randomly resampled the regions with
replacement in order to construct $N_\mathrm{b} = 300$ mock samples
\citep[e.g.,][]{norberg09}. For each bootstrap sample, $k$, constructed in this
way, we estimated $w_{\mathrm{p}}^k(r_{\mathrm{p}})$ using
Eqs.~\ref{eq:xi2d} and~\ref{eq:wprp} and constructed the bootstrap
covariance matrix via
\begin{equation}
  C = \frac{1}{N_\mathrm{b}-1} \sum_{k=1}^{N_\mathrm{b}}
  \left(
    w_{\mathrm{p},i}^k - \overline{w}_{\mathrm{p},i}
  \right)
  \left(
    w_{\mathrm{p},j}^k - \overline{w}_{\mathrm{p},j}
  \right),
  \label{eq:cov}
\end{equation}
where $i, j$ refer to bins of $r_{\mathrm{p}}$. The large-scale bias was
then estimated for each bootstrap sample using $\chi^2$ minimization, namely\
minimizing $\Delta^T C^{-1} \Delta$, where $\Delta$ is the difference between
the data and the model (Eq.~\ref{eq:bias}) and $C^{-1}$ is the inverse of the
covariance matrix.
For the projected two-point correlation function, the large-scale bias, and the typical DM halo mass, we
report the errors as the $16\%, 50\%$, and $84\%$ percentiles of the bootstrap
distribution based on $N_\mathrm{b} = 300$ samples.

\section{Results}%
\label{sec:results}

Here we describe the results of the AGN clustering analysis as a function of
the obscuration ${N_\mathrm{H}}$ and redshift. In particular, we divided the
AGN sample into two redshift bins:
$0.3 \le z < 1.1$ and $1.1 \le z < 3.0$,
which we refer to as the low and high redshift samples in the following (see
Table~\ref{tab:allevato2011}). The two bins are defined to
contain a comparable number of AGNs with $N =
\NumberRedshiftLowObscurationAll{} $ and $
\NumberRedshiftHighObscurationAll{}$
for the low and high redshift intervals. We note that changing the exact cut in
redshift by $\pm 0.2$ corresponds to a difference of ${\sim} 5-10\%$ in the
large-scale bias.

For each redshift bin we defined three AGN subsamples based on different cuts
in ${N_\mathrm{H}}$, which we refer to as unobscured (${N_\mathrm{H}} < 10^{22}
\,\mathrm{cm}^{-2}$), moderately obscured ($10^{22} \leq {N_\mathrm{H}} /
\,\mathrm{cm}^{-2} < 10^{23.5}$), and highly obscured (${N_\mathrm{H}} \geq
10^{23.5}$) sources. We also note that in the 2 Ms CDF-N and 7 Ms CDF-S,
\citet{li20} only report the X-ray spectral analysis for moderately to highly
obscured AGNs with ${N_\mathrm{H}} > 10^{23} \,\mathrm{cm}^{-2}$. A difference
of $\pm 0.2\,\mathrm{dex}$ in the ${N_\mathrm{H}}$ cut corresponds to a
difference of ${\sim} 5-10\%$ in the large-scale bias estimates and does not
change the overall trend of our results. It is worth noting that the
${N_\mathrm{H}}$ cut at $10^{23.5}\,\mathrm{cm}^{-2}$ (compared to the
${N_\mathrm{H}} >10^{24}\,\mathrm{cm}^{-2}$ cut that defines the CTK AGN
regime) is driven by the need to increase the statistics in each AGN
subsample. In fact, the highly obscured AGN bin is the most sensitive to the
exact choice of the limits in redshift and obscuration, as the number counts
decrease rapidly toward higher redshifts and higher obscuration.

\begin{figure*}[htbp]
  \centering
  \includegraphics[width=1.00\linewidth]{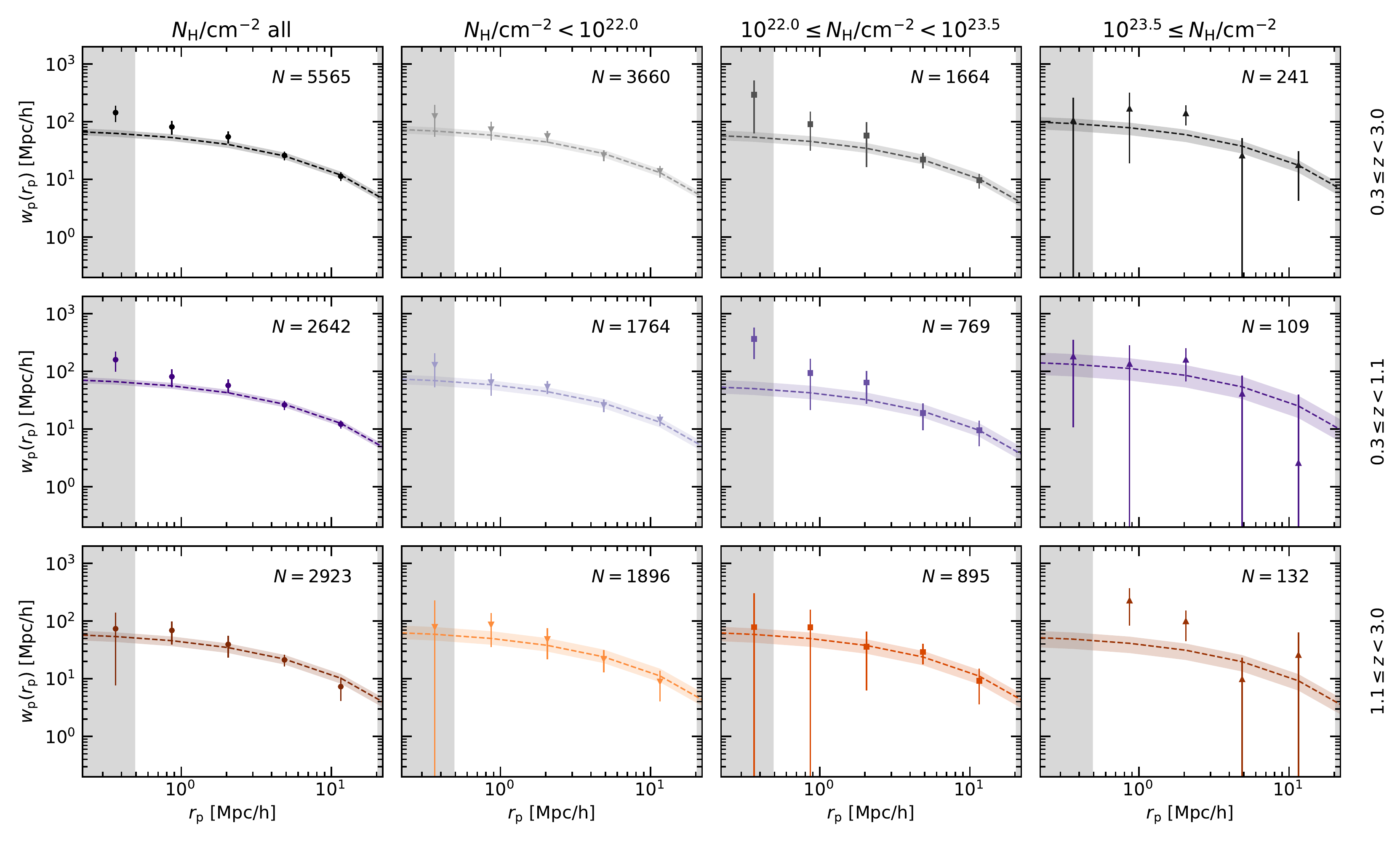}
  \caption{%
    Measured two-point correlation function in redshift (rows) and
    obscuration bins (columns). Each figure shows the
    $w_{\mathrm{p}}(r_{\mathrm{p}})$, with error bars corresponding to the $1\sigma$
    derived from bootstrap resampling. The best-fit weighted bias,
    $\overline{b}_{A11}$, derived through the projected DM correlation function
    via $w_\mathrm{p}(r_{\mathrm{p}}) = b^2 w_\mathrm{p,DM}(r_{\mathrm{p}},
    z=0)$ is shown as a dashed line with the same colors. The colored-shaded
    region corresponds to the $16\%$ and $84\%$ quantiles of the best-fit bias
    over the bootstrap samples. The shaded horizontal dark region corresponds
    to the $r_\mathrm{p}$ not included in the bias fitting.
  }%
  \label{fig:rp_wp_obscuration}
\end{figure*}

\begin{figure*}[htbp]
  \centering
  \includegraphics[width=\linewidth]{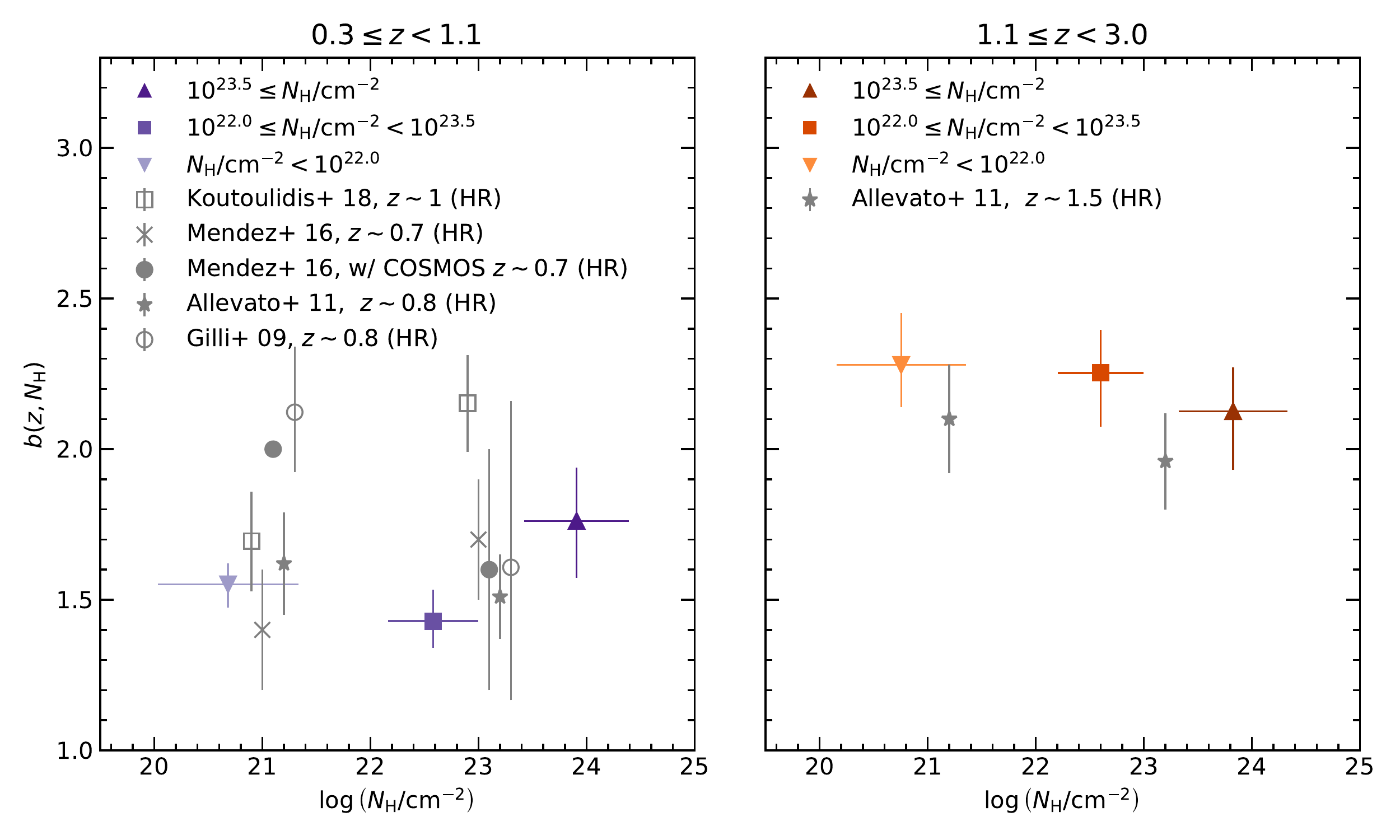}
  \caption{%
    Large-scale bias as a function of redshift (panels) and obscuration
    (markers in accordance with the legend). The vertical error bars correspond
    to the $16\%$\ and $84\%$ quantiles of the best-fit bias using bootstrap
    resampling. The horizontal error bars show the $1\sigma$ of the
    ${N_\mathrm{H}}$ distribution of the sample.  Measurements from X-ray AGN
    clustering in terms of the HR at comparable redshifts are
    shown as gray markers in accordance with the legend
    \citep[][]{gilli09,allevato11,mendez16,koutoulidis18}. Note that for
    plotting purposes we assume ${N_\mathrm{H}} = 10^{21} \,\mathrm{cm}^{-2}$
    (${N_\mathrm{H}} = 10^{23} \,\mathrm{cm}^{-2}$) for the soft (hard) samples
    and include a slight offset in ${N_\mathrm{H}}$ for visual clarity. With
    open markers we show the bias based on the correlation length measurement, $r_0$,
    assuming a power-law index $\gamma=1.8$, as reported by~\cite{gilli09},
     \cite{allevato11}, \cite{mendez16}, and \cite{koutoulidis18}.
  }%
  \label{fig:obscuration_bias}
\end{figure*}

\begin{figure*}[htbp]
  \centering
  \includegraphics[width=\linewidth]{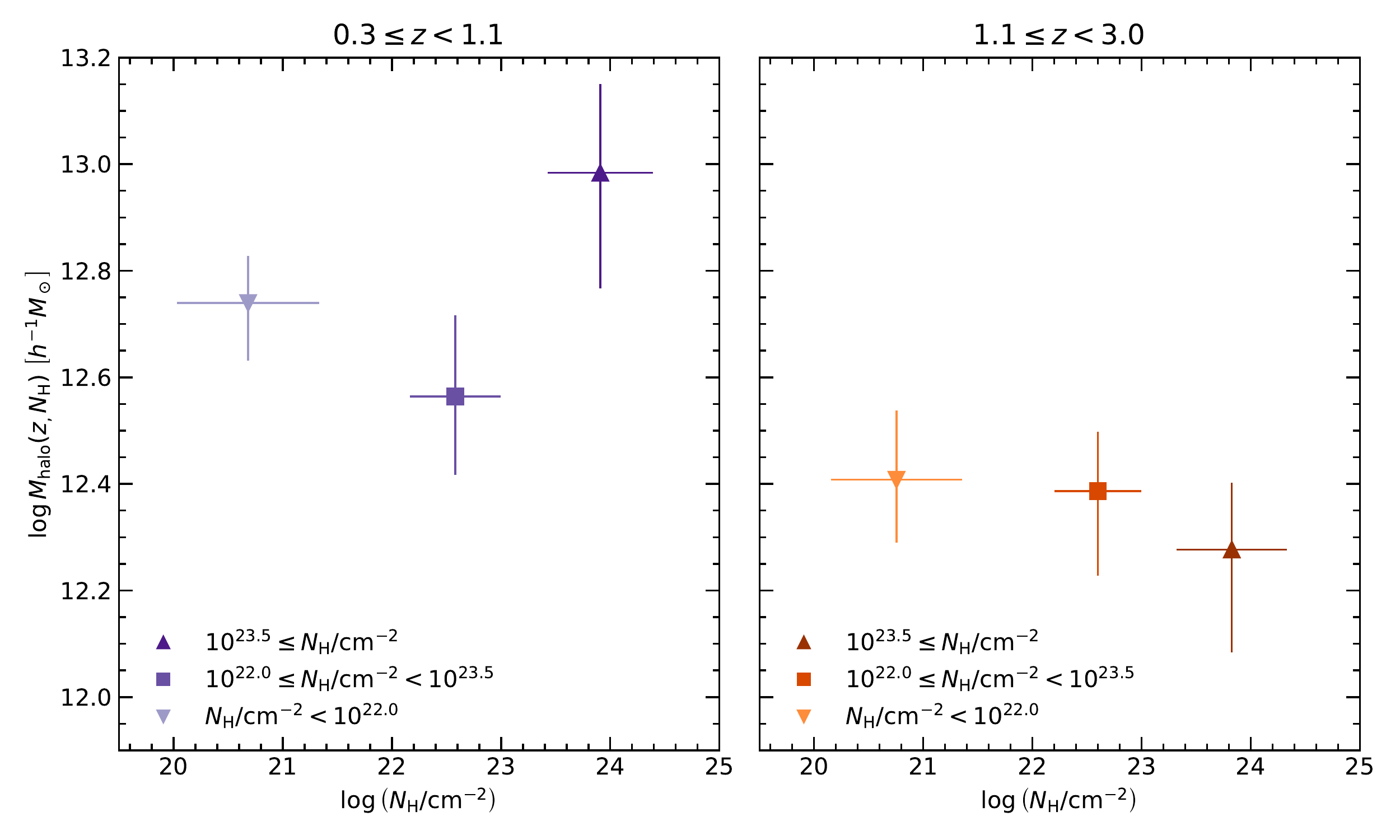}
  \caption{%
    Typical DM halo mass as a function of redshift. The colors and
    markers carry the same meaning as in Fig.~\ref{fig:obscuration_bias}.%
  }%
  \label{fig:obscuration_mhalo}
\end{figure*}

We show the estimated AGN projected two-point correlation function,
$w_{\mathrm{p}}(r_{\mathrm{p}})$, as a function of ${N_\mathrm{H}}$ for the
entire redshift range and for the low and high redshift intervals in
Fig.~\ref{fig:rp_wp_obscuration}. We then derived the large-scale bias,
$b(z,{N_\mathrm{H}})$ (Fig.~\ref{fig:obscuration_bias}), and the corresponding
typical DM halo mass (Fig.~\ref{fig:obscuration_mhalo}) at
different obscuration cuts and redshifts using the two different methodologies described in Sect.~\ref{sec:methodology}. We focus our discussion on the
results based on the methodology put forth by \citet[][]{allevato11}, and
provide the two-halo large-scale bias and the corresponding typical DM halo mass
for reference. We summarize all the estimates in Table~\ref{tab:allevato2011}.
It is worth noting that by modeling the large-scale clustering, only a
{typical} mass of the hosting halos can be derived; it may not be
representative of the entire hosting halo mass distribution of the considered
AGN sample. Only the modeling of the small-scale signal can provide this
information.

At high redshift we find no dependence of the AGN large-scale bias on the
level of obscuration of ${N_\mathrm{H}}$. In terms of the corresponding DM hosting
halos, our results show, for the first time, that unobscured and
moderately to highly obscured AGNs reside in the same environment at $z \sim 1.8$,
at any ${N_\mathrm{H}}$. At low redshift, $z \sim 0.7$, moderately obscured AGNs
are slightly less biased ($b = \BiasAllevatoRedshiftLowObscurationObscured{}$) than
highly obscured sources ($b = \BiasAllevatoRedshiftLowObscurationCTK{}$), but
the bias estimates are consistent within the error bars.

\begin{figure*}[htbp]
  \centering
  \includegraphics[width=.49\linewidth]{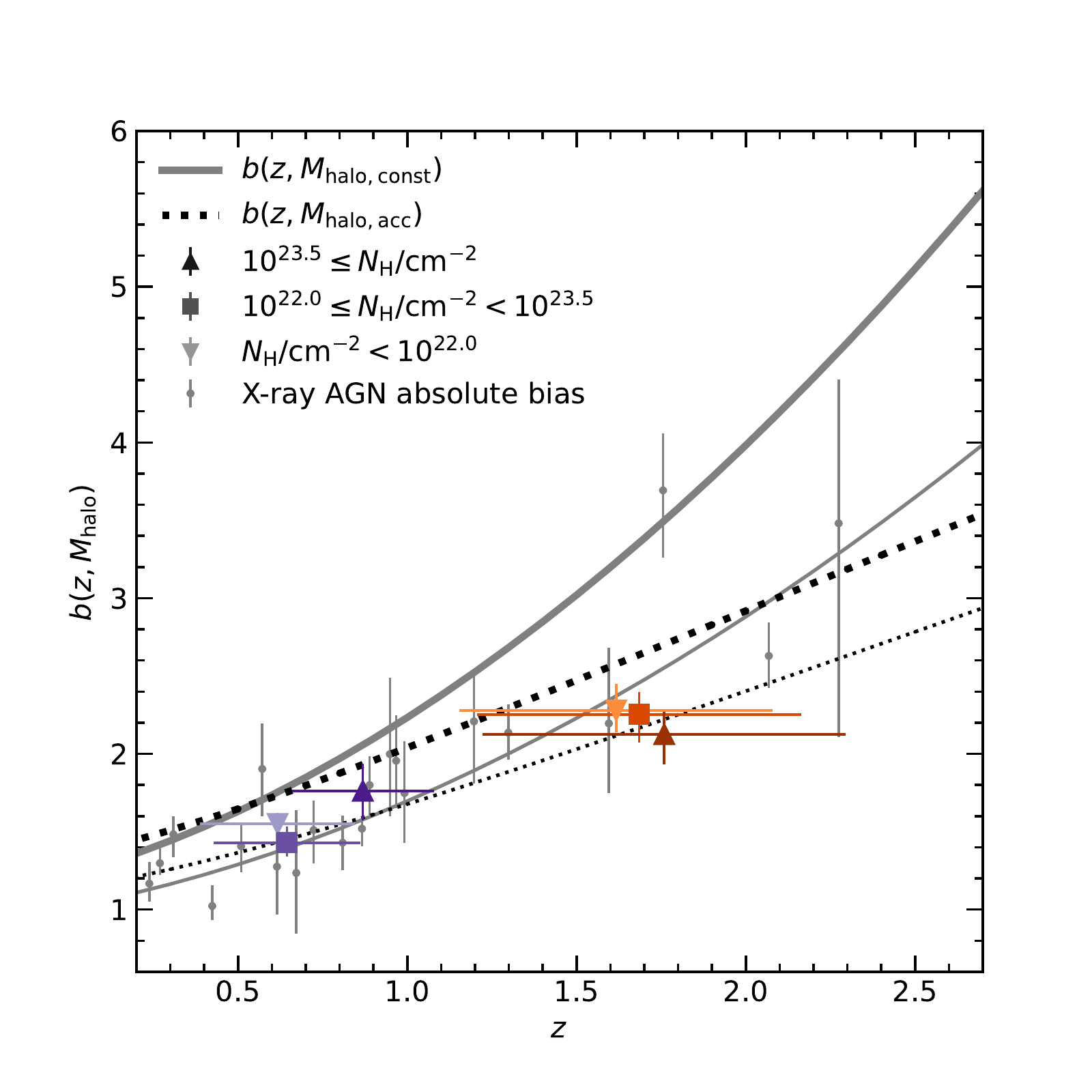}
  \includegraphics[width=.49\linewidth]{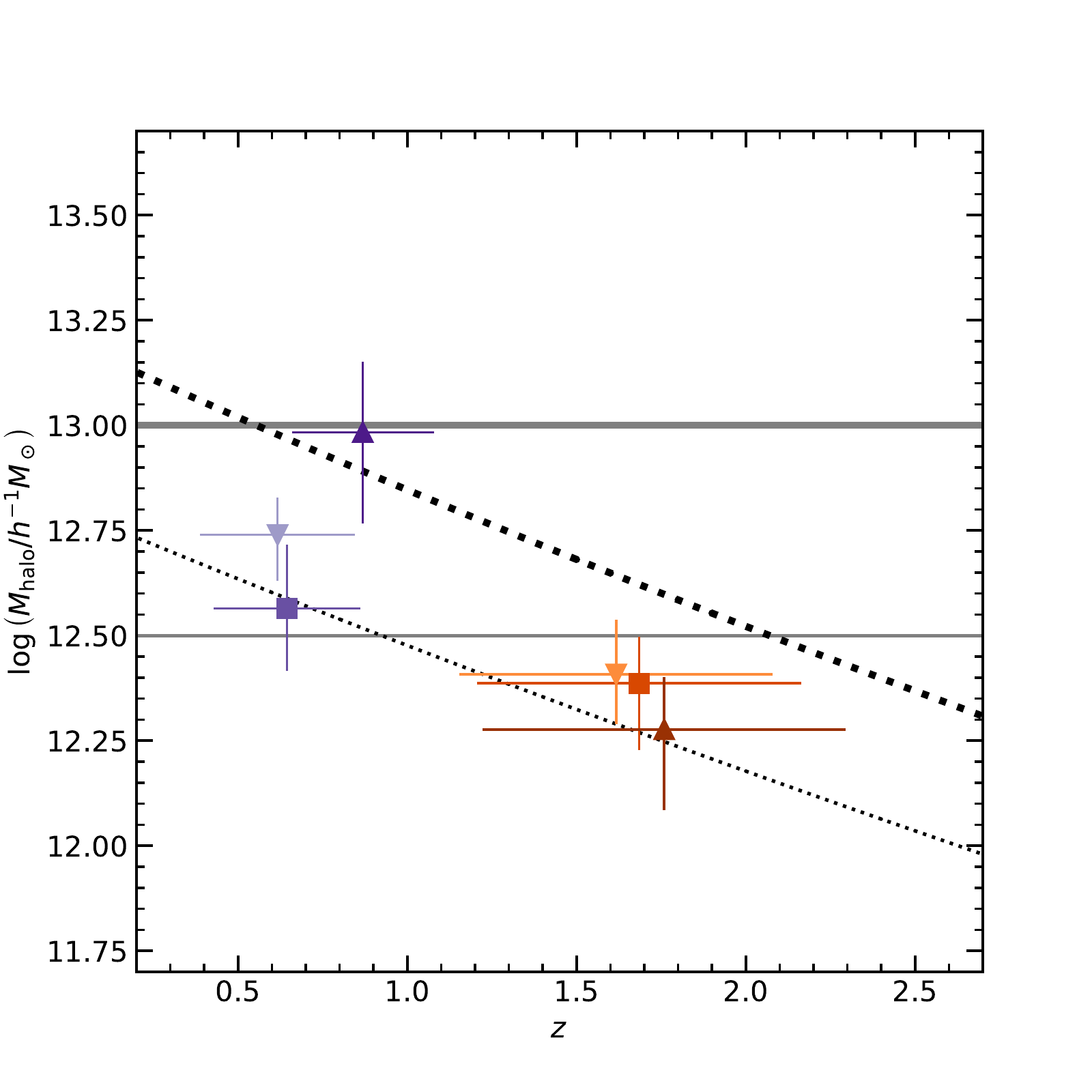}
  \caption{%
    Large-scale bias (left) and typical DM halo mass (right) as a function of
    redshift and AGN obscuration. The markers and colors correspond to the
    obscuration bins, in accordance with the legend. The horizontal error bars
    correspond to $1\sigma$ of the redshift distribution. The solid lines show
    the redshift evolution of the bias for a constant DM halo mass, $b(z,
    M_\mathrm{halo} = \mathrm{const}) $ and for DM halo masses $\log M_\mathrm{halo}
    = 12.5$\ and $ 13.0$ (in units of $\,h^{-1}M_\odot$) using the prescriptions of
    \citet{sheth01} and \citet{vandenbosch02} (labeled
    $M_\mathrm{halo,const}$). The dotted lines track the DM halo mass evolution
    through the accretion rates given by \citet[][]{fakhouri10} given a
    present-day DM halo mass of $\log M_\mathrm{halo} = 12.8, 13.2$ (in
    $\,h^{-1}M_\odot$, labeled $M_\mathrm{halo,acc}$). The gray markers show
    the X-ray-selected AGN bias measurements as reported by \citet[see the text
    for the details]{aird2021MNRAS.502.5962A}.
  }%
  \label{fig:redshift_bias_mhalo}
\end{figure*}

Figure~\ref{fig:redshift_bias_mhalo} shows the large-scale bias and the
corresponding typical mass of the hosting halos as a function of
redshift for all the AGN samples with different ${N_\mathrm{H}}$ cuts. We find
that the typical mass of the hosting halos decreases with redshift
irrespective of the obscuration cut. In detail, the
halo mass of unobscured (highly obscured) AGNs decreases with increasing
redshift, going from
$\log
{M_\mathrm{halo}}\,[h^{-1}\,\mathrm{M}_\odot] =
\LogMhaloAllevatoRedshiftLowObscurationUnobscured{}$ ($ =
\LogMhaloAllevatoRedshiftLowObscurationCTK{}$) at $z \sim 0.7$ to
$= \LogMhaloAllevatoRedshiftHighObscurationUnobscured{}$ ($ =
\LogMhaloAllevatoRedshiftHighObscurationCTK{}$) at $z \sim 1.8$. A similar
trend is observed for moderately obscured AGNs.

\subsection{Different combination of the input surveys}%
\label{sec:combinations}

\begin{figure*}[htbp]
  \centering
  \includegraphics[width=\linewidth]{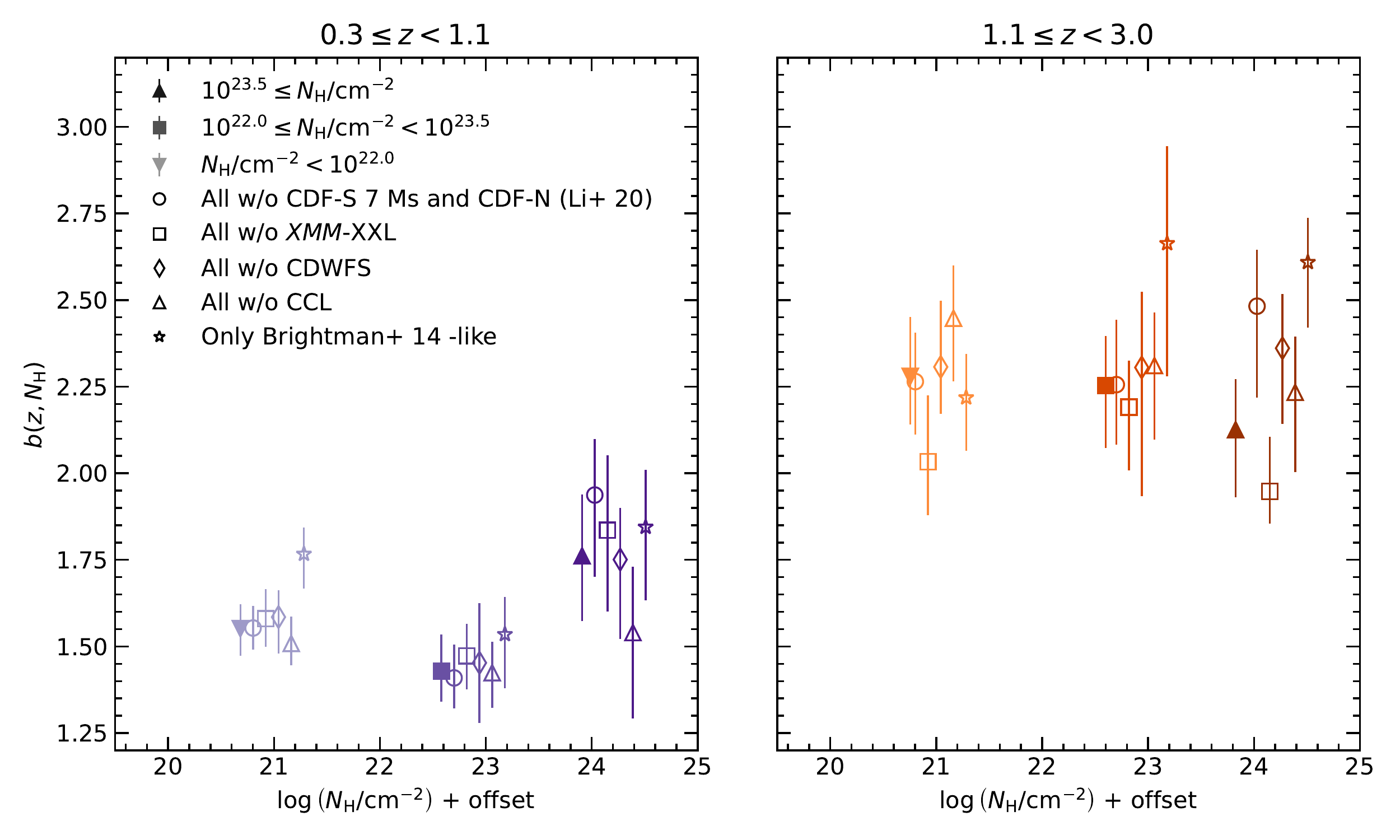}
  \caption{%
    Large-scale bias as a function of hydrogen column density, $N_H$, when
    excluding different surveys from the combined AGN sample. In each set, the
    leftmost point shows the original reported result when including all eight
    surveys in the analysis. The adjacent open symbols (offset slightly for
    visual clarity) correspond to when one or more surveys are left out of the
    combined sample. The circles correspond to when CDF-S 7Ms and CDF-N
    AGNs are excluded \citep[][]{li20}. Squares, diamonds, and triangles correspond to
    when AGNs from \emph{XMM}-XXL, CDWFS, and CCL are excluded,
    respectively. Finally, star symbols correspond to when we include only the surveys
    considered in \citet[][]{brightman14}, i.e., AEGIS-XD, 4 Ms CDF-S,
    C-COSMOS, and X-UDS \citep[][]{kocevski18}. See Sect.~\ref{sec:combinations}
    for details.
  }%
  \label{fig:combinations}
\end{figure*}

We further explored whether our results are affected by combining in the
analysis different surveys and/or ${N_\mathrm{H}}$
derivation methodologies. For this, we performed recalculations of the projected
two-point correlation function and the corresponding large-scale bias, leaving
out one or more surveys at a time. In detail, we considered the following cases,
removing from the analysis:
(\textit{i}) 2 Ms CDF-N and 7Ms CDF-S AGNs from the catalog of~\cite{li20}, which only
includes sources with log${N_\mathrm{H}} > 23.0$;
(\textit{ii}) \emph{XMM}-XXL AGNs (i.e.,\ we omit the only \emph{XMM}-\textit{Newton} survey
included in this work);
(\textit{iii}) CDWFS AGNs for which the obscuration ${N_\mathrm{H}}$ has been estimated
by using the HR instead of X-ray spectral analysis;
and (\textit{iv}) CCL AGNs, given the overdensities caused by large
structures present in the COSMOS field \citep[e.g.,][]{gilli09,mendez16}.
As final test, we included in the analysis only the AGN catalog presented in
\citet[][]{brightman14}, which includes the AEGIS-XD, 4 Ms CDF-S, C-COSMOS, and X-UDS
\citep[][]{kocevski18} fields and where the X-ray spectral analyses are all
performed using the same methodology. We summarize the results of these
tests in Fig.~\ref{fig:combinations}.

In summary, we find that our results in terms of large-scale bias as a function
of obscuration ${N_\mathrm{H}}$ do not depend on whether or not we include AGNs
with ${N_\mathrm{H}}$ derived from the HR. Similar results are
obtained when excluding from the analysis \textit{XMM}-XXL and CCL AGNs\@. Since
the statistics rapidly decrease when moving to higher redshifts and higher
${N_\mathrm{H}}$ cuts, the sample of highly obscured AGNs at high redshifts, $z
\sim 1.8$, is the most affected by the particular combination of surveys used.
In fact, as expected, the data points at $z \sim 1.8$ for highly obscured AGNs
show a larger scatter. Furthermore, a larger bias is found when only the AGN
catalogs described in \citet{brightman14} and \citet{kocevski18} are included in the
analysis.
\begin{table*}[htbp]
  \centering
  {\footnotesize
    \begin{tabular}{rrrrrrrrrrrrrr}
      \toprule
      \toprule
      $z_{\min}$ & $\overline{z}$ & $z_{\max}$ & $N_{\mathrm{H},\min}$ & $\overline{\log N_{\mathrm{H}}}$ & $N_{\mathrm{H},\max}$ & $\overline{\log L_{\mathrm{X}}}$ & $N$ & $\overline{b}_\mathrm{A11}$ & $\overline{z}_\mathrm{A11}$ & $\log M_\mathrm{A11}$ & $b_\mathrm{A11}$ & $b_\mathrm{2-halo}$ & $\log M_\mathrm{2-halo}$\\
      \midrule
      0.30 & 1.26 & 3.00 & $-        $ & $21.27_{-1.27}^{+1.47}$ & $-        $ & $43.99_{-0.83}^{+0.58}$ & $    5\,565$ & $ 1.20_{-0.08}^{+0.09}$ & $ 0.95$ & $12.60_{-0.07}^{+0.07}$ & $ 1.74_{-0.06}^{+0.06}$ & $ 2.19_{-0.15}^{+0.16}$ & $12.71_{-0.13}^{+0.11}$ \\
      0.30 & 1.23 & 3.00 & $-        $ & $20.73_{-0.73}^{+0.79}$ & $10^{22.0}$ & $43.98_{-0.85}^{+0.56}$ & $    3\,660$ & $ 1.26_{-0.09}^{+0.10}$ & $ 0.90$ & $12.66_{-0.08}^{+0.07}$ & $ 1.79_{-0.07}^{+0.07}$ & $ 2.30_{-0.18}^{+0.17}$ & $12.79_{-0.14}^{+0.12}$ \\
      0.30 & 1.28 & 3.00 & $10^{22.0}$ & $22.58_{-0.40}^{+0.53}$ & $10^{23.5}$ & $44.02_{-0.82}^{+0.57}$ & $    1\,664$ & $ 1.12_{-0.10}^{+0.12}$ & $ 1.05$ & $12.50_{-0.10}^{+0.11}$ & $ 1.65_{-0.07}^{+0.09}$ & $ 2.03_{-0.18}^{+0.22}$ & $12.57_{-0.18}^{+0.19}$ \\
      0.30 & 1.41 & 3.00 & $10^{23.5}$ & $23.84_{-0.28}^{+0.68}$ & $-        $ & $43.86_{-0.66}^{+0.86}$ & $       241$ & $ 1.46_{-0.20}^{+0.16}$ & $ 1.33$ & $12.70_{-0.14}^{+0.08}$ & $ 1.82_{-0.12}^{+0.09}$ & $ 2.66_{-0.37}^{+0.29}$ & $13.03_{-0.24}^{+0.15}$ \\
      \midrule
      0.30 & 0.69 & 1.10 & $-        $ & $21.40_{-1.17}^{+1.31}$ & $-        $ & $43.44_{-0.59}^{+0.57}$ & $    2\,642$ & $ 1.24_{-0.08}^{+0.07}$ & $ 0.64$ & $12.71_{-0.06}^{+0.07}$ & $ 1.53_{-0.04}^{+0.05}$ & $ 1.76_{-0.11}^{+0.11}$ & $12.92_{-0.12}^{+0.11}$ \\
      0.30 & 0.69 & 1.10 & $-        $ & $20.68_{-0.68}^{+0.96}$ & $10^{22.0}$ & $43.42_{-0.60}^{+0.58}$ & $    1\,764$ & $ 1.26_{-0.12}^{+0.12}$ & $ 0.62$ & $12.74_{-0.11}^{+0.09}$ & $ 1.55_{-0.08}^{+0.07}$ & $ 1.80_{-0.18}^{+0.16}$ & $12.96_{-0.20}^{+0.15}$ \\
      0.30 & 0.69 & 1.10 & $10^{22.0}$ & $22.58_{-0.42}^{+0.52}$ & $10^{23.5}$ & $43.48_{-0.57}^{+0.51}$ & $       769$ & $ 1.08_{-0.13}^{+0.16}$ & $ 0.64$ & $12.56_{-0.14}^{+0.16}$ & $ 1.43_{-0.09}^{+0.10}$ & $ 1.53_{-0.20}^{+0.23}$ & $12.64_{-0.31}^{+0.28}$ \\
      0.30 & 0.77 & 1.10 & $10^{23.5}$ & $23.91_{-0.33}^{+0.54}$ & $-        $ & $43.50_{-0.64}^{+0.81}$ & $       109$ & $ 1.75_{-0.38}^{+0.40}$ & $ 0.87$ & $12.98_{-0.21}^{+0.17}$ & $ 1.76_{-0.19}^{+0.18}$ & $ 2.48_{-0.58}^{+0.56}$ & $13.47_{-0.41}^{+0.28}$ \\
      \midrule
      1.10 & 1.77 & 3.00 & $-        $ & $21.16_{-1.16}^{+1.62}$ & $-        $ & $44.35_{-0.44}^{+0.41}$ & $    2\,923$ & $ 1.12_{-0.12}^{+0.09}$ & $ 1.64$ & $12.36_{-0.11}^{+0.08}$ & $ 2.22_{-0.12}^{+0.10}$ & $ 2.45_{-0.25}^{+0.22}$ & $12.42_{-0.20}^{+0.15}$ \\
      1.10 & 1.74 & 3.00 & $-        $ & $20.75_{-0.75}^{+0.60}$ & $10^{22.0}$ & $44.33_{-0.41}^{+0.40}$ & $    1\,896$ & $ 1.17_{-0.14}^{+0.18}$ & $ 1.62$ & $12.41_{-0.12}^{+0.13}$ & $ 2.28_{-0.14}^{+0.17}$ & $ 2.56_{-0.30}^{+0.40}$ & $12.50_{-0.22}^{+0.24}$ \\
      1.10 & 1.79 & 3.00 & $10^{22.0}$ & $22.60_{-0.39}^{+0.52}$ & $10^{23.5}$ & $44.40_{-0.44}^{+0.42}$ & $       895$ & $ 1.16_{-0.17}^{+0.15}$ & $ 1.68$ & $12.39_{-0.16}^{+0.11}$ & $ 2.25_{-0.18}^{+0.15}$ & $ 2.53_{-0.44}^{+0.36}$ & $12.48_{-0.36}^{+0.22}$ \\
      1.10 & 1.93 & 3.00 & $10^{23.5}$ & $23.83_{-0.29}^{+0.72}$ & $-        $ & $44.28_{-0.71}^{+0.58}$ & $       132$ & $ 1.06_{-0.19}^{+0.15}$ & $ 1.76$ & $12.28_{-0.20}^{+0.12}$ & $ 2.13_{-0.20}^{+0.14}$ & $ 2.31_{-0.49}^{+0.34}$ & $12.31_{-0.48}^{+0.25}$ \\
      \bottomrule
    \end{tabular}
  }
  \caption{%
    Number of X-ray AGNs divided into bins of redshift and obscuration,
    and the results of the large-scale bias model. Columns 1-8
    correspond to the minimum, mean, and maximum redshift, $z$;
    the minimum, median, and maximum AGN hydrogen column density,
    ${N_\mathrm{H}}$ (in $\mathrm{cm}^{-2}$); the median X-ray luminosity (in
    $\mathrm{erg}\,\mathrm{s}^{-1}\,\mathrm{cm}^{-2}$); and the number of AGNs. For
    ${N_\mathrm{H}}$ and ${L_\mathrm{X}}$ (large-scale bias and typical DM halo mass), the lower and upper limits correspond to 16\% and 84\%
    percentiles of the (bootstrap) distribution. Note that we consider
    $10^{20}\,\mathrm{cm}^{-2}$ as the lower limit for unobscured AGNs\@.
    Columns 9-12 correspond to the weighted large-scale bias, weighted
    mean redshift, typical DM halo mass, and the corresponding large-scale
    bias assuming the results from \citet{sheth01} and  \citet{vandenbosch02} and using the methodology
    set forth by \citet[][]{allevato11}. The last two columns show the
    best-fit two-halo bias as defined in Eq.~\ref{eq:bias} and the
    corresponding typical DM halo mass using
    the results from \citet{sheth01} and \citet{vandenbosch02}.
  }%
  \label{tab:allevato2011}
\end{table*}

\section{Discussion}

In this work we measure the clustering properties as a function of
obscuration, defined in terms of hydrogen column density (${N_\mathrm{H}}$) or
HR,
of one of the largest compilations of X-ray-selected AGNs from eight deep
\textit{XMM}/\textit{Chandra} and multiwavelength surveys. For the first time, in
addition to measuring the clustering of unobscured (${N_\mathrm{H}} <
10^{22}\,\mathrm{cm}^{-2}$) and moderately obscured AGNs ($10^{22} \leq
{N_\mathrm{H}} / \mathrm{cm}^{-2} < 10^{23.5}$), we also targeted highly
obscured AGNs ($\geq 10^{23.5}\,\mathrm{cm}^{-2}$). Here we discuss our results
in a larger context.

\subsection{Unobscured versus moderately obscured AGNs}

We find that the AGN large-scale bias is independent of the obscuration level of
${N_\mathrm{H}}$ at $z \sim 1.8$.\ A small deviation from this trend is
observed for highly obscured AGNs, which are slightly more biased than their less
obscured counterparts at $z \sim 0.7$ (see Fig.~\ref{fig:obscuration_bias}).
However, the bias estimates are consistent within the error bars.

Previous studies at different redshifts show controversial results, namely\
unobscured sources being more biased than moderately obscured ones
\citep{cappelluti10,allevato11,allevato14}, or no significant difference in the
large-scale clustering of the two populations
\citep{gilli09,ebrero09,coil09,krumpe12,mendez16,mountrichas12}. However, the
fact that in these works AGNs are classified as obscured or unobscured sources based on
different methods (e.g., the HR, WISE colors, and optical spectroscopy)
might bias the results since, for example,\ the optical type classification does not
perfectly match the X-ray classification \citep{merloni14}.

Our AGN sample and chosen methodology are similar to those in the recent work by
\citet[][]{koutoulidis18}. They measured the clustering of 736 (720) unobscured
(obscured) X-ray-selected AGNs in five deep \textit{Chandra} fields over $0.6 < z < 1.4$
and with mean ${L_\mathrm{X}} \sim 10^{43} \,\mathrm{erg}\,\mathrm{s}^{-1}$. At this
redshift, $z \sim 1$, they find obscured sources (${N_\mathrm{H}} > 10^{22}
\,\mathrm{cm}^{-2}$) classified via the HR to be slightly more clustered than
unobscured AGNs\@. However, their trend might be driven by highly obscured
AGNs, which we find to be slightly more clustered compared to moderately
obscured AGNs.

Similarly, the majority of infrared-selected AGN clustering studies find that
obscured sources are located in more massive halos than their unobscured counterparts
\citep{hickox11,donoso14,dipompeo14,dipompeo15,dipompeo16,dipompeo17}. However,
it is worth noting that these studies use different wavelengths to select
and classify AGNs\@. In fact, \citet{koutoulidis18} show that the optical/infrared
criterion fails to identify a pure AGN sample and that obscured quasars based
on optical/infrared colors may equally contain X-ray-unobscured and X-ray-obscured AGNs\@.
\citet{mountrichas21} infer that optical/mid-infrared color criteria are better
suited for infrared-selected AGNs and that their efficiency drops for the low to moderate
luminosity sources included in X-ray samples.

In this work, instead, the AGN classification is based on the column density,
$N_\mathrm{H}$, which is more robust in identifying obscured sources in samples
of moderate luminosity AGNs, compared to those based on HR and/or infrared/optical
criteria. Moreover, this analysis includes, for the first time, highly obscured
AGNs with $N_\mathrm{H} \geq 10^{23.5}\,\mathrm{cm}^{-2}$. Our results are in
disagreement with a picture in which obscured AGNs are located in denser
environments than unobscured sources
\citep{koutoulidis18,hickox11,donoso14,dipompeo16,dipompeo17}, which has been
interpreted in previous studies \citep{king10,hickox11} as BH mass
growth lagging behind that of the hosting halo.

On the contrary, our results suggest no significant dependence of the AGN
clustering properties on the obscuration, in agreement with AGN unified models,
in which whether the AGN is observed as obscured or unobscured depends only on
viewing angle, and in which, statistically, the halo-scale environments should be the same
for both populations. While we do find a slightly higher large-scale bias
for the highly obscured AGNs compared to moderately obscured AGNs at low z, it is
still consistent within the errors.

It is worth noting that orientation-based models are oversimplified. In fact,
the orientation with respect to a nuclear obscuring torus might not be the main
driver of the differences between obscured and unobscured AGNs at any redshift.
We know that the circumnuclear geometry is not the only factor, as there is a
dependence of the obscuration on the luminosity, Eddington ratio, and galaxy and BH
masses \citep[e.g.,][]{ricci2017Natur.549..488R,lanzuisi2017A&A...602A.123L}.
Moreover, the AGN obscuration might also occur on host galaxy scales and be
related to the overall galaxy evolution. Intrinsic differences in terms of, for example, the host galaxy stellar mass and BH mass between unobscured and
moderately to highly obscured AGNs may also exist. Despite all this, it is
noteworthy that our results suggest a negligible dependence of the AGN
large-scale bias on the obscuration $N_\mathrm{H}$.

On the other hand, according to the evolutionary models, different levels of
obscuration correspond to different stages of the growth of BHs and galaxy
evolution
\citep{%
  ciotti1997ApJ...487L.105C,%
  dimatteo2005Natur.433..604D,%
  gilli2007A&A...463...79G,%
  hopkins2006ApJS..163....1H,%
  lapi2006ApJ...650...42L,%
  bournaud2007ApJ...670..237B,%
  somerville2008MNRAS.391..481S,%
  treister2009ApJ...696..110T,%
  fanidakis2011MNRAS.410...53F%
}.
In AGN evolutionary models, obscured AGNs are an evolutionary phase of
merger-driven BH fueling, in which the AGN is first obscured, followed by an
unobscured phase after a gas blowout. However, while these models can explain
unobscured AGNs in more massive halos than moderately obscured sources \citep[or
vice versa; see, e.g.,][]{dipompeo17}, this key prediction might be at odds with
the lack of significant dependence on obscuration we find here.

Investigation of the clustering properties of larger samples of moderately and,
in particular, highly obscured AGNs matched in terms of stellar and BH mass
distributions across a wide range of redshift and luminosity would be
fundamental to better understanding obscured AGNs\@. Current and future surveys,
such as eROSITA \citep[][]{merloni2012arXiv1209.3114M}, will definitely increase the number of moderately to highly
obscured AGNs available for clustering studies.

\subsection{Small-scale clustering of obscured AGNs}

Alternatively, there are studies of \textit{Swift}-BAT
\citep[][]{baumgartner2013ApJS..207...19B}
X-ray AGNs in the local Universe,
suggesting that unobscured  and obscured AGNs (defined according to $N_H$) mainly differ in
terms of small-scale clustering (one-halo term) at $r_\mathrm{p} \lesssim 0.5-1.0
\,{h}^{-1}\mathrm{Mpc}$, which is due to the clustering of AGN pairs within the
same DM halos \citep{krumpe18,powell18}. The modeling of this
signal puts constraints on the halo occupation distribution, that is to say, on\ how AGNs
populate central and satellite halos.

We find that at small scales, moderately obscured sources might have a slightly
higher clustering signal than unobscured sources, at least at low redshifts of $z
\sim 0.7$ (see Fig.~\ref{fig:rp_wp_obscuration}). This result, in terms of the
one-halo term, might also reflect (or be due to) the observed difference in the
bias of moderately obscured compared to unobscured AGNs\@. Unfortunately, for
highly obscured sources, the one-halo term is measured with a large uncertainty.

Following \citet{powell18}, one possible explanation is that the number of AGNs
in satellite galaxies grows with a steeper slope for obscured AGNs than
for unobscured AGNs. This means that at a fixed halo mass, the probability of
obscured AGNs being in satellite galaxies is higher than that of unobscured sources.
In particular, \citet{powell18} and \citet{krumpe18} argue that the dominant triggering
mechanism of obscured AGNs in groups and clusters is not mergers (as the merging
cross section decreases in the high relative velocity encounters) but more
likely disturbances due to close encounters and/or secular processes.

Differences in halo concentration can also explain the different small-scale
clustering signal of the two populations, with obscured AGNs more concentrated
than their unobscured counterparts \citep{gatti16,powell18}. Highly concentrated
halos of a given mass would have a high concentration of satellite galaxies and
therefore have a higher probability of galaxy interactions, such as minor
mergers and encounters. For instance, \citet{jiang16} report significantly more
satellites around narrow-line Type 1 AGNs compared to broad-line Type 1 AGNs in
the SDSS at low redshifts, $z < 0.09$.

Future and current surveys will significantly increase the number of
moderately to heavily obscured AGNs, allowing us to measure the AGN clustering
signal with higher accuracy at small scales, and to extend the modeling of the
one-halo term to higher redshifts than previous studies of AGNs in the local
Universe.

\subsection{Redshift evolution}

In Fig.~\ref{fig:redshift_bias_mhalo} we show the redshift evolution of the AGN
bias and the corresponding typical DM halo mass for the different AGN
samples, as summarized in
Table~\ref{tab:allevato2011}.
In addition, we plot a compilation of previous X-ray AGN clustering results
from $L_\mathrm{X}$-limited samples with varying numbers of AGNs at $0.3
\lesssim z \lesssim 2$ as presented by \citet{aird2021MNRAS.502.5962A}. The
references are given in \citet[][Fig.~11]{aird2021MNRAS.502.5962A}.
For comparison, we also show the halo mass evolution tracks when using (\textit{i}) the
bias evolution assuming a constant halo mass (dashed line) as a function of
redshift \citep{sheth01,vandenbosch02} and (\textit{ii}) the halo mass accretion history
(dotted line) assuming the accretion rate of \citet{fakhouri10}, based on the
Millennium-II N-body simulation.

We measured, for the first time, the large-scale bias and the typical
mass of the hosting halos for highly obscured AGNs and find that these objects
reside in dense environments typical of galaxy groups at low ($z \sim 0.7$) and
high ($z \sim 1.8$) redshifts. In particular, we find that the redshift
evolution of the AGN bias of unobscured and moderately obscured AGNs follows a
passive evolution track, implying that for these AGNs the clustering is mainly
driven by the growth rate of the host halos and galaxies across cosmic
time. Following the mass accretion rate of \citet{fakhouri10}, this corresponds
to DM halos with $M_\mathrm{halo}(z = 1.8) \sim 10^{12.3} M_\odot/h$ that
evolves into $M_\mathrm{halo}(z = 0.7) \sim 10^{12.6} M_\odot/h$. A small
deviation is observed for highly obscured AGNs, but given the large error bars,
a passive evolution with redshift is still consistent with the data.

Determining whether the BH growth in highly obscured AGNs is driven by the host galaxy
evolution while in moderately obscured AGNs it is regulated by a constant DM halo
mass required to host the AGN activity requires further investigation. In
fact, studies connecting galaxy stellar and halo masses indicate that a similar
halo mass growth as a function of redshift corresponds to a fixed stellar mass of
$\sim 10^{11} M_\odot$ \citep[i.e.,\ above the knee of the stellar-to-halo mass
relation;][]{leauthaud12,shuntov2022A&A...664A..61S}. The similar
behavior for X-ray AGNs found here could suggest that AGN activity is more
closely linked to the host galaxy stellar mass rather than the DM halo. As
already discussed in previous sections, a study of moderately to highly
obscured AGNs with known host galaxy and BH properties is required.

\section{Conclusions}

In this work we have measured the clustering properties of unobscured, obscured,
and (for the first time) highly obscured AGNs by compiling one of the largest
available samples of X-ray-selected AGNs from deep \textit{XMM-Newton}/\textit{Chandra}
surveys with known hydrogen column densities, ${N_\mathrm{H}}$, derived from
X-ray spectral analysis and/or the X-ray HR. We divided the sample into
two redshift bin, $z \sim 0.7$ and $z \sim 1.8$, and analyzed the
clustering for unobscured ($N_\mathrm{H} < 10^{22}\,\mathrm{cm}^{-2}$),
moderately obscured ($10^{22} \leq N_\mathrm{H} < 10^{23.5}\,\mathrm{cm}^{-2}$),
and highly obscured ($N_\mathrm{H} \ge 10^{23.5}\,\mathrm{cm}^{-2}$) AGNs separately\@. We
summarize our findings as follows.

\begin{enumerate}[(1)]

  \item We find that, irrespective of obscuration, the typical DM
    halo mass of X-ray AGN samples is ${M_\mathrm{halo}} \sim
    10^{12.5-13}\,h^{-1}\,\mathrm{M}_\odot$, similar to group-scale
    environments and in line with previous works of X-ray-selected AGNs\@.

  \item
      At low ($z \sim 0.7$) and high ($z \sim 1.8$) redshifts, our results
      suggest an AGN large-scale bias independent of the obscuration
      $N_\mathrm{H}$. A slightly larger bias is measured for highly obscured
      AGNs at low z, but it is still consistent within the errors. Despite the fact that
      orientation-based models are oversimplified and that the obscuration
      depends on host galaxy and BH properties, it is noteworthy that our
      results suggest a negligible dependence of the AGN large-scale bias on
      the obscuration $N_\mathrm{H}$.%

  \item We estimated, for the first time, the large-scale bias and the typical
    mass of the hosting halos for highly obscured AGNs and find that these
    objects reside in DM halos with a typical mass $\log
    M_\mathrm{halo} [M_{\odot}/h] \sim
    \LogMhaloAllevatoRedshiftLowObscurationCTKApprox$
    ($\LogMhaloAllevatoRedshiftHighObscurationCTKApprox{}$) at a low (high)
    redshifts, $z \sim 0.7$ ($1.8$).

  \item We find that the redshift evolution of the AGN bias follows a passive
    evolution track irrespective of obscuration, implying that for these AGNs
    the clustering is mainly driven by the growth rate of the hosting halos and
    galaxies with time.

\end{enumerate}

Our work can be easily extended by including the future and ongoing X-ray
surveys with spectroscopic follow-ups. The increase in number counts will be
especially important for the CTK AGNs. Such surveys include eROSITA, which is
already underway and is expected to yield a sample of a few million AGNs up to
$z \sim 6$, and the future survey by Athena
\citep[e.g.][]{nandra2013arXiv1306.2307N} and/or AXIS
\citep[][]{mushotzky2019BAAS...51g.107M}, which are expected to detect
${\sim}20\,000$ AGNs at $z > 3$ and at different levels of obscuration.

\begin{acknowledgements}
  AV acknowledges support from the Vilho, Yrjö and Kalle Väisälä Foundation of
  the Finnish Academy of Science and Letters.
  VA acknowledges support from INAF-PRIN 1.05.01.85.08.
  FS acknowledges partial support from the European Union’s Horizon 2020
  research and innovation programme under the Marie Skłodowska-Curie grant
  agreement No. 860744.
\end{acknowledgements}

\bibliographystyle{aa}

\end{document}